\documentclass[11pt,a4paper]{article}
\usepackage{jheppub}

\usepackage{hyperref}

\usepackage{caption}

\usepackage{cancel}

\usepackage{amsthm} 
\theoremstyle{plain}
\newtheorem{thm}{Theorem} 

\theoremstyle{definition}
\newtheorem{defn}{Definition} 
\newtheorem{exmp}{Example} 
 
\newtheorem{prop}{Proposition}

\title{Volumes of Polytopes Without Triangulations}
\author[a]{Michael Enciso}

\affiliation[a]{Mani L. Bhaumik Institute for Theoretical Physics\\
Department of Physics and Astronomy\\
University of California at Los Angeles\\
Los Angeles, CA 90095, USA }

\emailAdd{menciso@physics.ucla.edu}

\abstract{The geometry of the dual amplituhedron is generally described in reference to a particular triangulation.  A given triangulation manifests only certain aspects of the underlying space while obscuring others, therefore understanding this geometry without reference to a particular triangulation is desirable.  In this note we introduce a new formalism for computing the volumes of general polytopes in any dimension.  We define new ``vertex objects'' and introduce a calculus for expressing volumes of polytopes in terms of them.  These expressions are unique, independent of any triangulation, manifestly depend only on the vertices of the underlying polytope, and can be used to easily derive identities amongst different triangulations.  As one application of this formalism, we obtain new expressions for the volume of the tree-level, $n$-point NMHV dual amplituhedron.
  }   
\begin{document}
\maketitle

\section{Introduction}
In recent years our understanding of scattering amplitudes in both gauge theory and gravity has grown immensely.  New mathematical structures as well as novel techniques for calculation have been uncovered, leading to new perspectives on the nature of amplitudes and streamlining 
many computations in comparison with
the textbook Feynman diagram approach (see the recent reviews~\cite{ElvangRev,DixonRev, HennRev} and
references therein).  One area in which striking progress has been made is the study of amplitudes in maximally supersymmetric gauge and gravity theories, due to their added computational simplicity~\cite{SimplestQFT}.

One of the major breakthroughs in the study of maximally
supersymmetric gauge theories is the discovery of the amplituhedron,
an object that encodes all tree-level amplitudes and loop-level
integrands in planar $\mathcal{N} = 4$ super-Yang--Mills theory
(sYM)~\cite{Amplituhedron,IntoAmplituhedron}.  In the case of tree-level amplitudes, the amplituhedron is a region of a positive Grassmannian which encodes the amplitude through a canonical volume form with logarithmic singularities on its boundary~\cite{Amplituhedron, BigGrassmannian,Postnikov}.  

Specializing to $M_n^{\text{NMHV}}$---the $n$-point tree-level NMHV amplitude in planar $\mathcal{N}=4$ sYM---the amplitude obtained in this way
is naturally interpreted as the volume of a polytope in the $\mathbb{CP}^4$ that is dual to the space in which the amplituhedron lives~\cite{NoteOnPolytopes,Amplituhedron}.  This interpretation is due to the fact that these amplitudes can be expressed (e.g. using BCFW recursion~\cite{BCFW}) as a sum of objects that are naturally viewed as volumes of four-dimensional simplices.  In particular, we have~\cite{NoteOnPolytopes}
\begin{equation} 
M_n^{NMHV}=\sum_{i,j=1}^n[*i (i+1) j (j+1)] \label{NMHVTree}
\end{equation} 
where 
\begin{equation}
[ijklm]\equiv \frac{1}{4!}\frac{\langle ijklm\rangle^4}{\langle ijklP_0\rangle \langle jklmP_0\rangle \langle klmiP_0\rangle \langle lmijP_0\rangle \langle mijkP_0\rangle}, 
\end{equation} 
$\langle ijklm\rangle\equiv \epsilon_{\alpha\beta\gamma\delta\sigma}Z_i^{\alpha}Z_j^{\beta}Z_k^{\gamma}Z_l^{\delta}Z_m^{\sigma},$ the $Z_i^{\alpha}$'s are (bosonified, super) momentum-twistors encoding the external kinematics~\cite{SpuriousPoles}, $Z_*^{\alpha}$ is a fixed reference twistor (of which $M_n^{\text{NMHV}}$ is independent), the sum on $i$ and $j$ is cyclic modulo $n,$ and
\begin{equation}P_0^{\alpha}=\begin{pmatrix}0 &\\ 0 &\\ 0 &\\ 0 &\\ 1\end{pmatrix}.\end{equation}
Each five-bracket $[ijklm]$ is viewed as the volume a four-dimensional simplex in $\mathbb{CP}^4$~\cite{NoteOnPolytopes}, and therefore these polytopes are understood primarily through particular triangulations.  The equivalence of two different sums of simplices (as obtained for example by performing two different BCFW shifts, or equivalently by making different choices for $Z_*^{\alpha}$ in (\ref{NMHVTree})) can then be interpreted as a result of using two different triangulations of the same underlying polytope.

For N$^k$MHV tree amplitudes with $k>1,$ BCFW recursion again expresses the amplitude as a sum of terms, with different BCFW shifts leading to different expressions of the amplitude.  One is therefore motivated to also view these sums as different triangulations of some underlying geometric entity, and this entity is referred to as the dual amplituhedron. For $k>1$ no clear geometric picture of this dual exists, though there are indications that such a picture should exist~\cite{PositiveAmplituhedron, TowardsVolume}.

The equality of different expressions for these amplitudes from different BCFW shifts has been understood using global residue theorems (GRTs) in the Grassmannian~\cite{BigGrassmannian, MasonSkinner, DualityForSMatrix}.   In particular, $n$-point N$^k$MHV tree amplitudes can be obtained via contour integrals in the Grassmannian $G(k,n),$ and the GRTs can be used to derive different expressions for the same amplitude.  It was this understanding that led to the discovery of the amplituhedron.  However, for $k>1$ the geometry of the dual amplituhedron remains obscured.

In this note we aim to gain a better understanding of the NMHV tree level amplituhedron geometry by expressing its volume in a way that makes no explicit reference to a particular triangulation.  In doing so, we make precise combinatorial sense of what an $n$-polytope in $\mathbb{CP}^n$ refers to, and introduce objects referred to as ``vertex objects'' that can be used to straightforwardly write down the volume of a polytope directly from its combinatorial data.  We refer to this procedure as the ``vertex calculus,'' as it results in an expression for the volume of a polytope that is unique, independent of any particular triangulation, and manifestly dependent only on the vertices of the underlying polytope.  These vertex objects also satisfy simple so-called ``cohomological identities'' that manifest the same identities that the GRTs do in the Grassmannian picture of Refs.~\cite{BigGrassmannian, MasonSkinner, DualityForSMatrix}.

The outline of this paper is as follows.  In section 2 we define ``vertex 2-polytopes'' (i.e., two-dimensional polytopes, or polygons) as well as the two-dimensional vertex objects, and introduce the two-dimensional vertex calculus.  In section 3 we briefly go through these steps again in three dimensions.  In section 4 we do the same in four dimensions, introducing ``vertex 4-polytopes,'' the four-dimensional vertex objects, and the four-dimensional vertex calculus.  We also apply this calculus to NMHV tree amplitudes in planar $\mathcal{N}=4$ sYM.  Only in section 2 do we provide proofs of all statements, and in sections 3 and 4 we leave these out as they are similar in nature to their two-dimensional analogues.

\section{Vertex 2-Polytopes}
\begin{figure}[t!]
\centering
\captionsetup{width=0.8\textwidth}
    \includegraphics[scale=0.3]{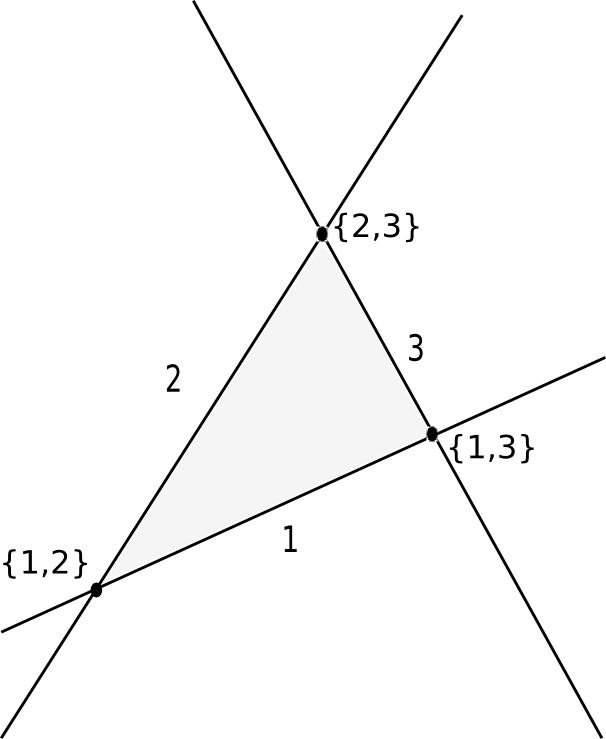}
    \caption{The depiction of a triangle whose edges are linearly embedded $\mathbb{CP}^1$'s in a $\mathbb{CP}^2.$} \label{fig:BasicTriangle}
\end{figure}

We begin by quickly recalling the basic facts about $\mathbb{CP}^n$ that we need before specializing in this and the next two sections, respectively, to the cases $n=2, 3,$ and 4.   Any $Z^{\alpha}\in \mathbb{CP}^n$ (with $n+1$ homogeneous coordinates labeled by $\alpha=0, 1, ..., n$) determines a unique linearly embedded $\mathbb{CP}^{n-1}$ in a dual $\mathbb{CP}^n$, whose elements $W_{\alpha}$ have $n+1$ homogeneous coordinates labeled by lower indices.  This correspondence between points and hyperplanes is realized via the usual linear homogenous pairing between variables and their duals: $W\cdot Z\equiv W_{\alpha}Z^{\alpha}=0$.  We assume an understanding of the intersection properties of $k$-planes in projective space, and details can be found in, for example, the appendix of Ref.~\cite{CP5Calculus}.  We also note that we use the notation $Z^{\alpha}$ to denote points in projective space since $Z$ is the standard letter used to denote (bosonified, super) momentum-twistors in $\mathbb{CP}^4,$ however unless otherwise stated the $Z^{\alpha}$'s will have no relation to kinematical data and should be viewed simply as generic points in $\mathbb{CP}^n.$\\
\indent Given three generic $Z_{1}^{\alpha}, Z_2^{\alpha}, Z_3^{\alpha}\in \mathbb{CP}^2$, we obtain three distinct lines in the dual $\mathbb{CP}^2$, each pair intersecting in a unique point.  As is convention, we refer to linearly embedded $\mathbb{CP}^1$'s (in some background $\mathbb{CP}^n$) as lines, $\mathbb{CP}^2$'s as planes, $\mathbb{CP}^3$'s as hyperplanes, and so on, even though topologically a line and a linearly embedded $\mathbb{CP}^1$ are very different.  Indeed, this distinction is precisely what we aim to make sense of, and the key realization is that even though lines and planes are topologically distinct from $\mathbb{CP}^1$'s and $\mathbb{CP}^2$'s, the (combinatorial) intersection structure of these objects behave identically in the complex-projective setting as they do in the real-projective setting.   Figure \ref{fig:BasicTriangle} depicts this scenario, with each line being labelled by its defining $Z_i^{\alpha}$ in the dual space, and the vertices of the ``triangle'' being labelled by the pair of lines whose intersection defines it.

\begin{figure}[t!]
\centering
\captionsetup{width=0.8\textwidth}
    \includegraphics[scale=0.3]{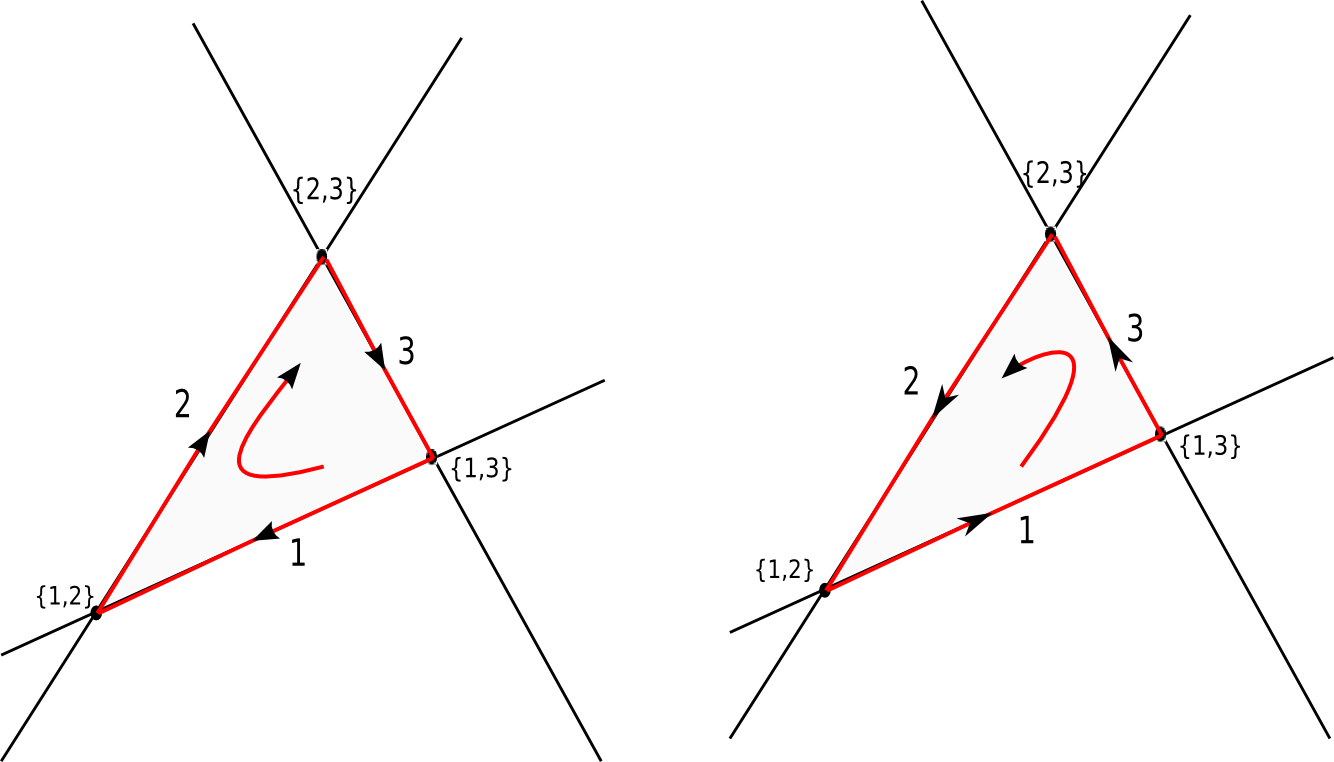}
    \caption{The two different orientations of a triangle.}\label{fig:DoubleTriangle}
\end{figure}

We aim to make rigorous the interpretation of Figure \ref{fig:BasicTriangle} as a triangle, even though it does not properly reflect the topological qualities of the region defined by $Z_1^{\alpha}, Z_2^{\alpha}, Z_3^{\alpha}$ (seeing as, for example, the ``edges'' are topologically spheres, as they are linearly embedded $\mathbb{CP}^1$'s).  Despite this, Figure \ref{fig:BasicTriangle} does correctly reflect how the projective lines intersect, and this is enough to define a meaningful notion of polygon.  In particular, we choose to view polygons as abstract instructions for moving from one vertex to another along a well-defined edge.  In Figure \ref{fig:BasicTriangle}, it is clear that we can move from vertex $\{1,2\}$ to $\{2,3\}$ along the line 2.  The fact that the line 2 is topologically an $S^2$ (being a linearly embedded $\mathbb{CP}^1$) is irrelevant, and this shows the utility of considering only combinatorial data.  Indeed, $\mathbb{CP}^2$ is four-real-dimensional and so we must be precise in what we mean by a two-polytope in this space.\\
\indent The instructions ``move from $\{1,2\}$ to $\{2,3\}$ along the line 2'' can be unambiguously denoted by $\{1,2\}\rightarrow \{2,3\}$ where it is understood that we move along the line whose label is common to the two vertices.  We denote the instructions $\{1,2\} \rightarrow \{2,3\}$ by $[1(2)3]$, making manifest both the starting and ending vertices as well as the line that joins them (in this case, 2).  It is then clear how Figure \ref{fig:BasicTriangle} can be viewed as being one of two oppositely oriented triangles: one orientation corresponds to the instructions 
\begin{equation}
\{1,2\} \rightarrow \{2,3\} \rightarrow \{3,1\} \rightarrow \{1,2\} \Big(=[1(2)3]+[2(3)1]+[3(1)2]\Big)\label{eq:Triangle1}
\end{equation}
while the opposite orientation corresponds to the instructions
\begin{equation}
\{2,1\} \rightarrow \{1,3\} \rightarrow \{3,2\} \rightarrow \{2,1\} \Big(=[2(1)3]+[1(3)2]+[3(2)1]\Big) \label{eq:Triangle2}.
\end{equation}
These two different scenarios are depicted respectively on the left and right of Figure \ref{fig:DoubleTriangle}.

\begin{figure}[t!]
\centering
\captionsetup{width=0.8\textwidth}
    \includegraphics[scale=0.45]{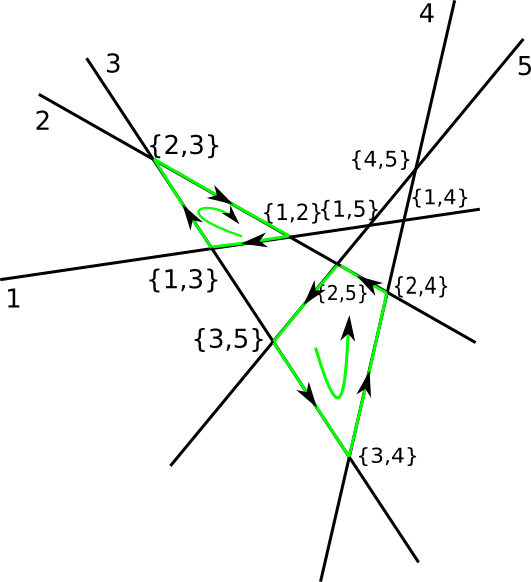} 
        \caption{The polygon corresponding to the list $l=(3,2,1,3,4,2,5).$}\label{fig:ArbitraryPolygon}
\end{figure}

We use these ideas to define arbitrary polygons by noting that the quality of the ``boundary of the boundary'' vanishing is simply that the instructions in (\ref{eq:Triangle1}) and (\ref{eq:Triangle2}) end where they began.  Thus, given any set of $N$ distinct elements $Z_1^{\alpha}, ..., Z_{N}^{\alpha}\in \mathbb{CP}^2$ defining a set of $N$ distinct lines in the dual space, we can view any set of instructions
\begin{equation}
\{i_1,i_2\}\rightarrow \{i_2, i_3\}\rightarrow \cdots \rightarrow \{i_m, i_1\}\rightarrow \{i_1, i_2\},
\end{equation}
with each\footnote{From here on we simply let $\{1, ..., N\}$ denote the set $\{Z_1^{\alpha}, ..., Z_N^{\alpha}\}\in \mathbb{CP}^n$ where $n=2$ here and later in higher dimensions $n$ will be obvious from context.} $i_k\in \{1, ..., N\}$, as defining a polygon.  We say that such a set of abstract instructions is given by a \textbf{cyclic list} $l=(i_1, ..., i_m)$, where $m$ simply denotes the length of the list.\\  
\indent For example, with boundaries $\{1,2,3\}$, the triangle on the left in Figure \ref{fig:DoubleTriangle} is defined by the list $l=(1,2,3)$, and with boundaries $\{1, 2, 3, 4, 5\}$, the ``polygon'' defined by the list $l=(3,2,1,3,4,2,5)$ is depicted in Figure \ref{fig:ArbitraryPolygon}.  We emphasize that we still have yet to define what we mean by ``polygon,'' but it is clear from Figure \ref{fig:ArbitraryPolygon} that we are building up to a definition that is not restricted to what we might want to call convex, or even connected polygons.\\
\indent Many different lists correspond to what we will want to call the same polygon.  For example, the cyclic permutation of any list defines equivalent instructions.  However, the situation is more subtle than that, as for example the lists $l'=(321342425)$ and $l''=(3425321)$ also define the same polygon as that depicted in Figure \ref{fig:ArbitraryPolygon}. In order to introduce an equivalence of lists it is most useful to use the $[\cdot(\cdot)\cdot]$ notation and introduce some more formal machinery.

We can place (and implicitly already have placed in, e.g., (\ref{eq:Triangle1}) and (\ref{eq:Triangle2})) an additive structure on the formal objects $[i(j)k]$ in a natural way: $-[k(j)i]\equiv [i(j)k]$ and $[i(j)k]+[k(j)l]\equiv[i(j)l]$.  These definitions denote respectively the oriented nature of each edge and the fact that walking from point $A$ to point $B$ then from point $B$ to point $C$ (all along the same edge) is equivalent to walking from point $A$ to point $C$.  An \textbf{edge set} $E$ is any formal sum of $[i(j)k]$ objects (called \textbf{oriented edges}), and if we define the boundary operator $\partial [i(j)k]\equiv \{j,k\}-\{i,j\}$, then we see that ``the boundary of the boundary'' of the ``polygon'' defined by $E$ vanishes if and only if $\partial E=0$.\\
\indent Given a cyclic list $l=(i_1, ..., i_n)$, we define the \textbf{edge set derived from $l$}, $E(l)$, as
\begin{equation}
E(l)\equiv \sum_{j=1}^n[i_{(j-1)}(i_j)i_{(j+1)}],
\end{equation}
where the sum on $j$ is cyclic in the sense that $i_{(n+1)}=i_1$.  Thus, for $l=(1,2,3)$, which determines the triangle on the left in Figure \ref{fig:DoubleTriangle}, we have
\begin{equation}
E(l=(123))=[1(2)3]+[2(3)1]+[3(1)2],
\end{equation}
as expected from (\ref{eq:Triangle1}), and for $l=(3213425)$ defining the polygon in Figure \ref{fig:ArbitraryPolygon}, we have
\begin{align}
E(l=(3213425))=&[3(2)1]+[2(1)3]+[1(3)4]+[3(4)2]\\ \nonumber
&+[4(2)5]+[2(5)3]+[5(3)2],
\end{align}
which are both readily checked to coincide with their respective figures.  The equivalence relation amongst lists is now clear: we say that $l_1\sim l_2$ if $E(l_1)=E(l_2)$.  One can readily check that the lists $l,l',$ and $l''$ given above corresponding to Figure \ref{fig:ArbitraryPolygon} are all equivalent.\\
\indent It can be shown that, for any edge set $E$, $\partial E=0 \Leftrightarrow E=E(l)$ for some cyclic list $l$.  We therefore make the following definition of, for lack of a better term\footnote{We are hesitant to use the term ``combinatorial polytope,'' as this terminology is already used in the mathematical literature and restricts itself to convex polytopes, which is a restriction we do not wish or need to impose.}, a ``vertex 2-polytope.''  It allows for a very general type of polygon, including disconnected polygons and possibly polygons with various components ``wrapped around'' many times.
\begin{defn}
A \textbf{vertex 2-polytope} $P$ is equivalent to the following data.\\
\indent i) A set $S=\{1, ..., N\}$ of $N$ distinct boundaries (lines) in the above sense.  \\
\indent ii) A cyclic list $l=(i_1, ..., i_n)$ with each $i_k\in S$.\\
\end{defn}

\subsection{Volumes of Vertex 2-Polytopes}
In line with Ref.~\cite{NoteOnPolytopes}, we say that the area $A(l)=A((1,2,3))$ of the triangle defined by the list $l=(1,2,3)$ is
\begin{equation}
A((1,2,3))=\frac{1}{2}\frac{\langle 123\rangle^2}{\langle 12P_0\rangle \langle 23P_0\rangle \langle 31P_0\rangle}\equiv [123] \label{eq:TriangleArea}
\end{equation}
where $P_0^{\alpha}\in \mathbb{CP}^2$ is a fixed reference boundary and $\langle ijk\rangle\equiv \epsilon_{\alpha\beta\gamma}Z_i^{\alpha}Z_j^{\beta}Z_k^{\gamma}$.  This is a natural generalization of the usual area of a triangle in real affine space, and we refer the reader to Ref.~\cite{NoteOnPolytopes} for details of this definition of complex-projective area.  We note that $[123]$ is projectively well-defined in $Z_1^{\alpha},$ $Z_2^{\alpha},$ and $Z_3^{\alpha},$ and its non-trivial weight in $P_0^{\alpha}$ defines the scale of the area. To generalize this notion of area to arbitrary polygons, we must note that there is a natural way to add two cyclic lists to get a cyclic list corresponding to the ``superposition'' (including orientation) of the two respective polygons.  Namely, if $l_1=(i_1, ..., i_n)$ and $l_2=(j_1, ..., j_m)$ are two cyclic lists, then $l=(i_1, ..., i_n, i_1, j_1, ..., j_m, j_1)$ is defined to be the sum of $l_1$ and $l_2$, and it can be readily checked that 
\begin{equation}
E(l)=E(l_1)+E(l_2).
\end{equation}
It is also clear that the equivalence class of the sum of two lists depends only on the equivalence class of the two summands.  For example, the list $l=(3,2,1,3,4,2,5)$ is in the same equivalence class as $l_1+l_2$ where $l_1=(3,2,1)$ and $l_2=(3,4,2,5)$, by noting that $l_1+l_2\equiv (3,2,1,3,3,4,2,5,3)\sim (3,2,1,3,4,2,5)$.\\
\indent The generalization of (\ref{eq:TriangleArea}) to arbitrary cyclic lists is then clear once we impose the reasonable condition that 
\begin{equation}
E(l_1)+E(l_2)=E(l)\Rightarrow A(l_1)+A(l_2)=A(l),
\end{equation}
meaning that the area of the ``superposition'' of two vertex 2-polytopes should be the sum of the areas of the individual polytopes.  We then have that if $l=(i_1, ..., i_n)$ is a cyclic list, the area $A(l)$ of the corresponding 2-polytope is 
\begin{equation}
A(l)=\sum_{k=1}^n[i_ki_{k+1}B] \label{eq:ArbitraryPolygonArea}
\end{equation} 
where $B^{\alpha}\in \mathbb{CP}^2$ is a fixed reference boundary, $A(l)$ is independent of our choice of $B^{\alpha}$, and the sum on $k$ is cyclic in the above sense.  This can all be seen by noting that 
\begin{equation}
l=(i_1, ...,i_n)\sim(i_1, i_2, B)+(i_2, i_3, B)+...+(i_{(n-1)},i_n,B)+(i_n,i_1,B)
\end{equation}
for any $B^{\alpha}$.  \\
\indent Equation (\ref{eq:ArbitraryPolygonArea}) gives the area of an arbitrary polygon in terms of a \textbf{particular} triangulation.  There are many different triangulations---i.e., sums of $[ijk]$ objects---that give an equivalent area for a given polygon, and identifying equivalent triangulations is often difficult.  The formalism introduced in the next subsection will be oblivious to any choice of triangulation, and we simply use (\ref{eq:ArbitraryPolygonArea}) to verify that our expression of the area of a polygon in terms of the vertex objects that we define is indeed valid.

\subsection{Two-Dimensional Vertex Calculus}
The key objects in the new formalism are ``vertex objects'' of the form $F_{i_1, ..., i_d}$ where $d$ is the dimension of the polytope under consideration.  These are the objects that we will use to express volumes in a triangulation-independent manner, and which we claim can be viewed as ``atomic'' objects when computing volumes of polytopes, replacing the $[a_1...a_{d+1}]$ objects---which are volumes of simplices---in this role.  The vertex objects come from certain cohomological considerations that we will only allude to in this work, leaving the details for a future note.  In this paper we will define the vertex objects as particular sums of the $[a_1...a_{d+1}]$ objects.  Some of the definitions we make here may appear arbitrary at first sight.  This is a consequence of our defining the \textbf{more} fundamental vertex objects in terms of the \textbf{less} fundamental $[a_1...a_{d+1}]$ objects (instead of vice-versa), and doing so without appealing to the cohomological motivations for the definitions.  The formalism relies on the fact that the vertex objects satisfy certain ``cohomological'' identities.  These allow us to take the expression for a volume in terms of vertex objects and algebraically obtain any triangulation.  We now turn to defining the vertex objects in two dimensions and seeing this cohomological identity explicitly.\\
\indent   We begin by quoting Ref.~\cite{NoteOnPolytopes} and stating without proof the following:
\begin{equation}
[124]+[234]+[314]=[123]\label{eq:FundamentalIdentity},
\end{equation}
for any boundaries $Z_1^{\alpha},Z_2^{\alpha}, Z_3^{\alpha}, Z_4^{\alpha}\in \mathbb{CP}^2$.  As in Ref.~\cite{SpuriousPoles} we interpret this result as the vanishing sum of the areas of four overlapping oriented triangles.  Proving this identity algebraically is non-trivial, thus making the geometric picture simpler.\\  \indent We now define the objects $V[ij][kl]$ as follows:
\begin{equation}
V[12][34]\equiv [123]-[124],\label{eq:2DQuadrilateral}
\end{equation}
and it is clear that each $V[\cdot\cdot][\cdot\cdot]$ is antisymmetric under swapping the entries in a particular $[\cdot\cdot]$.  Using (\ref{eq:FundamentalIdentity}), one can show that each $V[\cdot\cdot][\cdot\cdot]$ is also antisymmetric under swapping the $[\cdot\cdot]$ brackets themselves, i.e., $V[12][34]=-V[34][12]$.  We note that by choosing $B^{\alpha}$ to be $1, 2, 3$, or $4$ in  (\ref{eq:ArbitraryPolygonArea}), one can show that $V[12][34]=A(l=(1,4,2,3))$, so that each $V[\cdot\cdot][\cdot\cdot]$ corresponds to the area of a ``quadrilateral''.  It can also be shown using (\ref{eq:FundamentalIdentity}) that for any $Z_1^{\alpha}, Z_2^{\alpha},Z_3^{\alpha},Z_4^{\alpha},Q^{\alpha}$, 
\begin{equation}
V[12][3Q]+V[12][Q4]=V[12][34]\label{eq:2DUsefulRelation}.
\end{equation}
This result reflects the fact that ``slicing'' a quadrilateral with a line allows one to express the volume of the original quadrilateral as the sum of the two resulting quadrilaterals.  Thus the $V[\cdot\cdot][\cdot\cdot]$ objects have algebraic properties that make calculating with them easy, despite the fact that their ``inner workings'' in terms of the $Z_i^{\alpha}$ variables are rather complex.  In higher dimensions the inner workings of the analogous objects are more complex, but their algebraic properties are just as simple.  This is the first departure from the standard way of computing these volumes---areas of quadrilaterals are viewed as more fundamental than areas of triangles (though still less fundamental than the vertex objects that we soon define).\\
\indent We now define the main objects of our calculus and establish the cohomological identity.
\begin{defn} \label{def:2Df}
Let $S=\{1, ..., N\}$ be a set of $N$ distinct boundaries, and let $Q^{\alpha}\in \mathbb{CP}^2$ be a fixed reference boundary.  For each $i,j\in S$, we define the ``vertex objects''
\begin{equation}
F_{ij}=\sum_{k\neq i,j}^NV[ij][kQ]\label{eq:2Df}.
\end{equation}
\end{defn}
Due to the antisymmetry in $i$ and $j$ of each $F_{ij}$, Definition \ref{def:2Df} implicitly defines $N\choose 2$ non-trivial functions.  We also note that while each $F_{ij}$ depends on our choice of $Q^{\alpha}$, we suppress this dependence in our notation because we will soon see that the sums of vertex objects that we will be interested in are independent of this choice.  We now present the cohomological identity in two dimensions.
\begin{prop}\label{prop:main2DProp}
Let $S=\{1, ..., N\}$ be a set of $N$ distinct boundaries and let $\{F_{ij}\}$ be as in Definition \ref{def:2Df}.  Then for any $i,j,k\in S$ we have the following ``cohomological identity'': 
\begin{equation}
\rho_{[i}F_{jk]}\equiv F_{ij}+F_{jk}+F_{ki}=2[ijk].\label{eq:main2DIdentity}
\end{equation}
\end{prop}
Before proving this identity, we note a few of its important characteristics.  First, it is very similar to the $\check{\text{C}}$ech cocycle condition as written in Ref.~\cite{Huggett} for the first cohomology group (hence referring to it as the cohomological identity). This is no accident, but its detailed explanation will be given in a later note.  Secondly, we note that this result can be summarized by saying that a particular sum of quadrilaterals results in a two-fold covering of the triangle $[ijk]$.   We now give the proof of this result.\\
\textbf{Proof of Proposition \ref{prop:main2DProp}:}
\begin{align*}
F_{ij}+F_{jk}+F_{ki}&=\sum_{l\neq i,j}^N V[ij][lQ]+\sum_{l\neq j,k}^N V[jk][lQ]+\sum_{l\neq k,i}^N V[ki][lQ]\\
&=\sum_{l\neq i,j,k}^N(V[ij][lQ]+V[jk][lQ]+V[ki][lQ])\\
&\ \ \ \ +V[ij][kQ]+V[jk][iQ]-V[jQ][ki]\\
&=2[ijk]
\end{align*}
$\Box$\\
\\
The last equality comes from using (\ref{eq:2DUsefulRelation}) in the sum and expanding the definition of the $V[\cdot\cdot][\cdot\cdot]$'s in the remaining three terms.  Each individual $F_{ij}$ depends on our choice of reference boundary $Q^{\alpha}$ used to define it in (\ref{eq:2Df}), however the sums of the vertex objects that we will be interested in are independent of that choice.  Namely, we have the following.
\begin{prop}\label{prop:2DIndependenceOfQ}
Let $S=\{1, ..., N\}$ be a set of $N$ distinct boundaries, let $Q^{\alpha}$ and $Q'^{\alpha}$ be two reference boundaries, let $\{F_{ij}\}$ be as in Definition \ref{def:2Df} using $Q^{\alpha}$, and let $\{F'_{ij}\}$ be as in Definition \ref{def:2Df} using $Q'^{\alpha}$. Let $l=(i_1, ..., i_n)$ be a cyclic list with each $i_k\in S$, let $A=\sum_{k=1}^nF_{i_ki_{(k+1)}}$, and let $A'=\sum_{k=1}^nF'_{i_ki_{(k+1)}}$ where the sums on $k$ are cyclic.  Then $A=A'$.
\end{prop}

\textbf{Proof of Proposition \ref{prop:2DIndependenceOfQ}:}
\begin{align*}
A-A'&=\sum_{k=1}^n(F_{i_ki_{k+1}}-F'_{i_ki_{k+1}})\\
&=\sum_{k=1}^n\sum_{l\neq i_k, i_{k+1}}(V[i_ki_{k+1}][lQ]-V[i_ki_{k+1}][lQ'])\\
&=\sum_{k=1}^n\sum_{l\neq i_k, i_{k+1}}V[i_ki_{k+1}][Q'Q]\\
&=(N-2)\sum_{k=1}^nV[i_ki_{k+1}][Q'Q]\\
&=0.\ \ \ \ \ \ \ \ \ \ \ \ \ \ \ \ \ \ \ \ \ \ \ \ \ \ \Box
\end{align*}

We now give the statement of the two-dimensional calculus.
\begin{thm}\label{thm:Main2DThm}
Let $S=\{1, ..., N\}$ be a set of $N$ boundaries, and let $l=(i_1,...,i_n)$ be a cyclic list taking values in $S$.  Let $A(l)$ be the area of the vertex 2-polytope defined by $l$, as in (\ref{eq:ArbitraryPolygonArea}), with the reference boundary $B^{\alpha}$.  Let $S'=S\cup \{B^{\alpha}\}$ be the set of boundaries $\{1, ..., N, B\}$, and let $\{F_{ij}\}$ be defined with respect to $S'$ as in Definition \ref{def:2Df}.  Then $A(l)=\frac{1}{2}\sum_{k=1}^nF_{i_ki_{k+1}}$.  In other words, the volume of the polygon defined by $l$ is equal to $\frac{1}{2}A$ as defined in Proposition \ref{prop:2DIndependenceOfQ}.
\end{thm}

\textbf{Proof of Theorem \ref{thm:Main2DThm}:}
\begin{align*}
\frac{1}{2}\sum_{k=1}^nF_{i_ki_{k+1}}&=\frac{1}{2}\sum_{k=1}^n(F_{i_ki_{k+1}}+F_{i_{k+1}B}+F_{Bi_{k+1}})\\
&=\frac{1}{2}\sum_{k=1}^n(F_{i_ki_{k+1}}+F_{i_{k+1}B}+F_{Bi_k})\\
&=\frac{1}{2}\sum_{k=1}^n2[i_ki_{k+1}B]\\
&=\sum_{k=1}^n[i_ki_{k+1}B]=A(l)
\end{align*}
where in the first equality we simply added $0=F_{i_{k+1}B}+F_{Bi_{k+1}}$ to each term in the sum.  In the second equality we took advantage of the fact that the sum on $k$ is cyclic to relabel the index on the third term, and in the third equality we used (\ref{eq:main2DIdentity}). $\Box$
\\
\\
\indent  The two-dimensional vertex calculus for computing the area of a vertex 2-polytope is therefore as follows.  Take the cyclic list $l=(i_1, ..., i_n)$ which defines the polytope, and go down the list adding up $F_{i_ki_{k+1}}$ at each step---there is no need to think about any triangulation.   In practice, we never have to make reference to the actual structure of the $F_{ij}$'s because we can just use (\ref{eq:main2DIdentity}) to simplify our resulting sum of $F_{ij}$'s into a sum of areas of triangles, if we wish.  By making different choices of simplification, we can recover any triangulation.  Thus, this calculus does more than recover the particular triangulation used in (\ref{eq:ArbitraryPolygonArea}), and Theorem \ref{thm:Main2DThm} says that \textbf{any} triangulation obtained from this sum of $F_{ij}$'s will correspond to the correct area.  We also note that the sum $\sum_{k}F_{i_ki_{(k+1)}}$ is a sum over vertices, due to our definition of 2-polytopes in terms of cyclic lists, and if a vertex is ``spurious,'' as for example the $\{3,4\}$ vertex in the list $(1,2,3,4,3)$, then the corresponding $F_{...}$'s immediately cancel.  Thus, the vertex objects that remain in the sum label ``physical,'' i.e. ``genuine'' vertices via their subscripts.  Example \ref{ex:Quadrilateral} below illustrates some of the utility of the two-dimensional calculus.\\
\indent We quickly note that in Theorem \ref{thm:Main2DThm} we introduced the set of boundaries $S'=\{1, ..., N,B\}$ solely to make the expression $F_{iB}$ well-defined.  In practice, however, we only need to use $F_{ij}$ objects whose subscripts take values in $S=\{1, ..., N\}$.  We can extend this set of boundaries however we wish since Proposition \ref{prop:main2DProp}, which is the main result for the calculus, is independent of the set of boundaries we choose.  Equivalently, we could also choose $B^{\alpha}$ to be within the set $S$ itself, since $A(l)$ does not depend on our choice of $B^{\alpha}$.  The example in Appendix \ref{app:2DCalculus} shows this fact in practice, as well as gives some geometric insight into this calculus.
\begin{exmp} \label{ex:Quadrilateral}
Consider the cyclic list $l=(1,4,2,3)$, denoting the polygon depicted in Figure \ref{fig:Quadrilateral1}.
\begin{figure}[t!] 
\captionsetup{width=0.8\textwidth}
\centering
    \includegraphics[scale=0.4]{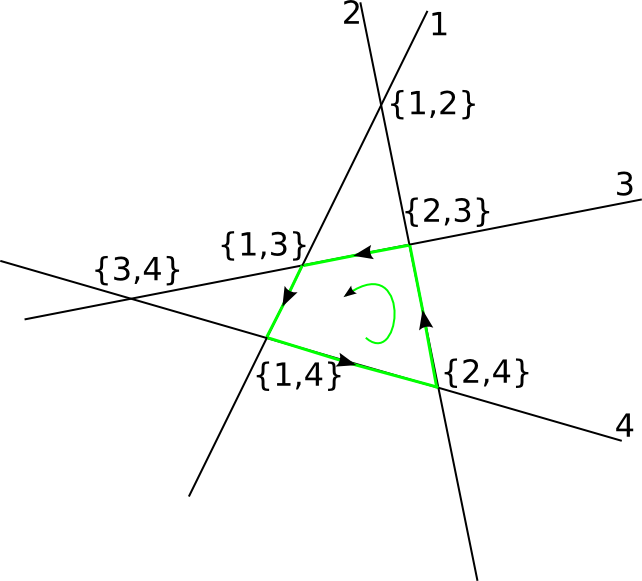} 
        \caption{The quadrilateral corresponding to the list $l=(1,4,2,3).$}\label{fig:Quadrilateral1}
\end{figure} 
Here, $N=4$ and so we construct the  $4\choose 2$ objects $\{F_{ij}\}$. Theorem \ref{thm:Main2DThm} tells us that
\begin{equation}
A(l)=\frac{1}{2}(F_{14}+F_{42}+F_{23}+F_{31})\label{eq:QuadrilateralInCalculus}.
\end{equation}
By simplifying this expression in two different ways using (\ref{eq:main2DIdentity}), we get
\begin{equation}
A(l)=[123]-[124]\label{eq:Triangulation1},
\end{equation}
as well as 
\begin{equation}
A(l)=[314]+[423]\label{eq:Triangulation2},
\end{equation}
which we know from (\ref{eq:FundamentalIdentity}) to be equal.  We have therefore recovered two different triangulations for the same 2-polytope.  Suppose however that we were given the expression 
\begin{equation}
(I)\equiv [145]+[425]+[236]+[316]+[156]+[265]\label{eq:BigTriangulation}
\end{equation}
for some boundaries $\{1, ..., 6\}$.  Using (\ref{eq:main2DIdentity}) in reverse, we can expand out each $[ijk]$ in terms of $F_{ij}$'s and make the obvious cancellations due to the antisymmetry of the vertex objects to get
\begin{align}
(I)=&\frac{1}{2}(F_{14}+\cancel{F_{45}}+\cancel{F_{51}}+F_{42}+\cancel{F_{25}}+\cancel{F_{54}}\\ \nonumber
&+F_{23}+\cancel{F_{36}}+\cancel{F_{62}}+F_{31}+\cancel{F_{16}}+\cancel{F_{63}}\\ \nonumber
&+\cancel{F_{15}}+\cancel{F_{56}}+\cancel{F_{61}}+\cancel{F_{26}}+\cancel{F_{65}}+\cancel{F_{52}})\\ \nonumber
=&\frac{1}{2}(F_{14}+F_{42}+F_{23}+F_{31})=A(l).
\end{align}
Thus, $(I)$ is seen to be both independent of the boundaries $5$ and $6$, as well as equal to $A((1,4,2,3))$, both of which are non-obvious when presented with (\ref{eq:BigTriangulation}).  To go the other direction and recover the triangulation $(I)$ from (\ref{eq:QuadrilateralInCalculus}) involves repeatedly adding zero in the form $F_{ij}+F_{ji}$ in a straightforward way.\\
\indent In the triangulations (\ref{eq:Triangulation1}), (\ref{eq:Triangulation2}), and (\ref{eq:BigTriangulation}), we have introduced spurious vertices into our triangulations.  For example, in order to move from (\ref{eq:QuadrilateralInCalculus}) to (\ref{eq:Triangulation1}) we had to add and subtract $F_{12}$, corresponding to the vertex $\{1,2\}$ which is not a part of the underlying polygon.  A physically interesting class of triangulations are those that do not introduce any such spurious vertices, as these give expressions of amplitudes without introducing spurious non-local poles~\cite{NoteOnPolytopes}.  The vertex calculus can straightforwardly recover these as well.  

\begin{figure}[t!] 
\captionsetup{width=0.8\textwidth}
\centering
    \includegraphics[scale=0.3]{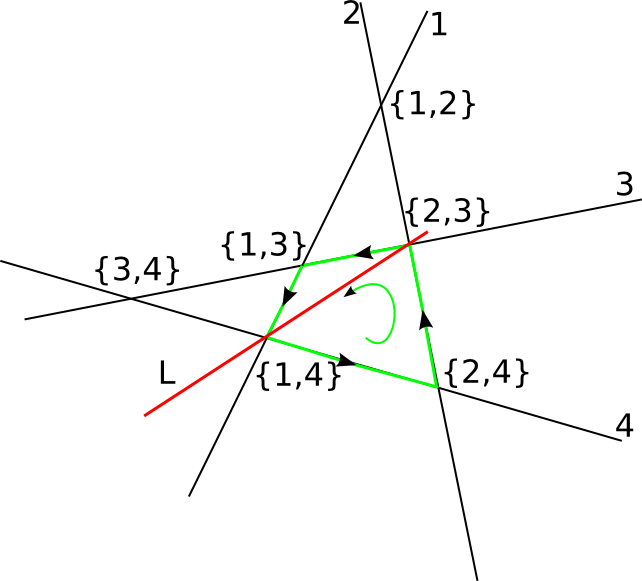} 
        \caption{A triangulation of the quadrilateral corresponding to the list $l=(1,4,2,3)$ that does not introduce any spurious vertices.}\label{fig:Quadrilateral2}
\end{figure}

For example, let us define $L$ to be the unique line connecting the vertex $\{1,4\}$ to the vertex $\{2,3\}$, as in Figure \ref{fig:Quadrilateral2}.  Then the areas $[14L]$ and $[23L]$ of the two triangles defined by these three lines both vanish, since the condition that $L$ intersects the vertex $\{i,j\}$ is that $\langle ijL\rangle=0.$  We therefore have 
\begin{equation}
F_{14}+F_{4L}+F_{L1}=2[14L]=0
\end{equation}  
and
\begin{equation}
F_{23}+F_{3L}+F_{L2}=2[23L]=0.
\end{equation}  
In particular, $F_{14}=F_{1L}+F_{L4}$ and $F_{23}=F_{2L}+F_{L3}$, so that we can rewrite (\ref{eq:QuadrilateralInCalculus}) as
\begin{equation}
A(l)=\frac{1}{2}(F_{1L}+F_{L4}+F_{42}+F_{2L}+F_{L3}+F_{31}) \label{eq:QuadrilateralInCalculus2}
\end{equation}
We note that all of the vertices labelled by the subscripts of the $F_{...}$'s in (\ref{eq:QuadrilateralInCalculus2}) are still physical (i.e., not spurious) vertices.  We can now use (\ref{eq:main2DIdentity}) (without introducing any new vertices) to find
\begin{equation}
A(l)=[L31]+[L42],
\end{equation}
as expected from Figure \ref{fig:Quadrilateral2}.\\
\indent Thus, any triangulation of $A(l)$ can be recovered from (\ref{eq:QuadrilateralInCalculus}) by adding zero and/or introducing new boundaries in various ways, and then simplifying the resulting sum using (\ref{eq:main2DIdentity}) in various ways.  The equality of any two triangulations can be straightforwardly checked.  It is also clear that we can recover the cyclic list itself from \textbf{any} triangulation of the underlying polygon.  Namely, given any triangulation, we can express it in terms of the vertex objects using (\ref{eq:main2DIdentity}).  We will then always be left with the expression (\ref{eq:QuadrilateralInCalculus}), which is readily seen to be obtained from the list $(1,4,2,3)$.  This process carries over directly for any list $l=(i_1, ..., i_n)$.  
\end{exmp}
To summarize, the vertex calculus expresses the area of any polygon in terms of objects whose subscripts label the physical vertices, giving an expression that is independent of any triangulation.  From this expression one can algorithmically obtain any triangulation by adding in spurious vertices and/or boundaries and using the cohomological identity (\ref{eq:main2DIdentity}).  We now turn our attention to developing the analogous formalism in higher dimensions.

\section{Vertex 3-Polytopes}
\begin{figure}[t!] 
\centering
\captionsetup{width=0.8\textwidth}
    \includegraphics[scale=0.3]{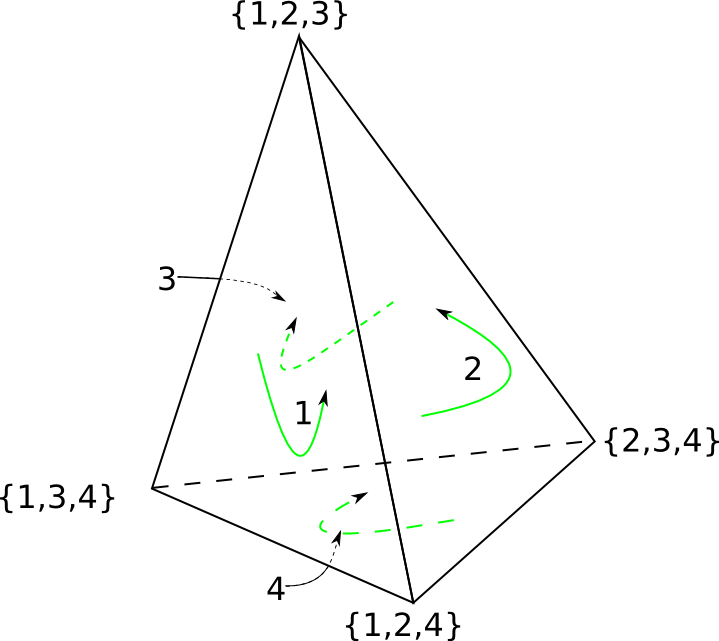} 
        \caption{A depiction of a three-simplex bounded by four $\mathbb{CP}^2$'s, and described by a cyclic list on each face.} \label{fig:3Simplex}
\end{figure}

We first seek to make precise what we mean by ``3-polytope'' in terms of cyclic lists, as these are what our three-dimensional vertex calculus will be based on. To do this, we view a 3-polytope as being a set of 2-polytopes ``glued together'' in such a way as to make the ``boundary of the boundary'' of the polytope vanish.\\
\indent We consider a set $S=\{1, ..., N\}$ of $N$ distinct boundaries $Z_{i}^{\alpha}\in \mathbb{CP}^3$, $1\leq i\leq N$, where now each $Z_i^{\alpha}$ determines a 2-plane---or linearly embedded $\mathbb{CP}^2$---in the dual $\mathbb{CP}^3.$  The intersection of any two distinct planes gives a line, and the intersection of any three gives a point.  Accordingly, an \textbf{oriented edge} is now of the form $[i(jk)l]$ with $i,j,k,l\in S$, corresponding to the instructions $\{i,j,k\} \rightarrow \{j,k,l\}$, meaning ``go from vertex $\{i,j,k\}$ to $\{j,k,l\}$ along the line defined by the intersection of the planes whose labels are common to the two vertices (in this case the line $j\cap k$)''.  The same additive structure can be placed on these formal objects, where now we can only add two oriented edges when both of their parenthetical entries are the same.   Thus, for example, $[1(23)4]+[1(34)5]$ is fully simplified whereas $[1(23)4]+[4(23)5]=[1(23)5]$.   To see how cyclic lists come in, we motivate our discussion by considering Figure \ref{fig:3Simplex}, which depicts the intersection structure of four boundaries $\{1, 2, 3, 4\}$ with a particular orientation.  We clearly want to view this as an oriented 3-simplex, but care is needed since each ``face'' of this object is really a linearly embedded $\mathbb{CP}^2.$

We can describe this object using four cyclic lists, one for each face.  Additionally, since the 2-polytope sitting on the $i^{th}$ face will, by definition, have the entry $i$ in each vertex, we need to change the instructions that our cyclic lists determine.  Namely, with $N$ boundaries, we get $N$ cyclic lists $\{l_i=(j_{i1}, ..., j_{in_i})\}_{1\leq i\leq N}$ with each $j_{ik}\in S$ and where $n_i$ is simply the length of the $i^{th}$ list.  Each list $l_i$---defining the 2-polytope on the $i^{th}$ face---defines the edge set $E_i\equiv E(l_i)$ as follows:
\begin{equation}
E(l_i)\equiv \sum_{k=1}^{n_i}[j_{i(k-1)}(ij_{ik})j_{i(k+1)}],
\end{equation}
with the sum on $k$ cyclic.  Thus, the four lists corresponding to Figure \ref{fig:3Simplex} are 
\begin{align}\label{eq:3SimplexLists} 
l_1=(2,3,4)\\ \nonumber 
l_2=(1,4,3)\\ \nonumber
l_3=(1,2,4)\\ \nonumber
l_4=(1,3,2),
\end{align}
and the edge set $E_1$, for example, is
\begin{equation}
E_1=[2(13)4]+[3(14)2]+[4(12)3],
\end{equation}
which is readily seen to agree with what we want to interpret as the ``triangle'' sitting on the boundary $1$.  We obtained the lists in (\ref{eq:3SimplexLists}) by orienting one of the boundary 2-faces and then orienting the rest in such a way that the ``boundary of the boundary'' vanished.  Thus these four lists are not independent, but rather satisfy certain ``gluing'' constraints.  To obtain these constraints more generally, we introduce the notion of an edge set $E_{i;s}$ which is the restriction of $E_i$ to the boundary $s\in S$ as follows:
\begin{equation}
E_{i;s}\equiv \sum_{j_{ik}=s}^{n_i}[j_{i(k-1)}(ij_{ik})j_{i(k+1)}]=\sum_{j_{ik}=s}^{n_i}[j_{i(k-1)}(is)j_{i(k+1)}].
\end{equation}  
Thus, for example, with respect to the lists in (\ref{eq:3SimplexLists}) we have $E_{1;2}=[4(12)3]$.  Since the ``boundary of the boundary'' of a 3-polytope should be the sum of its oriented edges, a vertex 3-polytope must be defined by lists $\{l_i\}$ such that $\sum_{i=1
}^NE_i=0$.  However, since edges in $E_{i;k}$ can only be cancelled by other edges in $E_{i;k}$ or by edges in $E_{k;i},$ and since we have
\begin{equation}
\sum_{i=1}^NE_i=\sum_{i}^N\sum_{k\neq i}E_{i;k}=\frac{1}{2}\sum_{i}^N\sum_{k\neq i}(E_{i;k}+E_{k;i}),
\end{equation}
we are motivated to make the following definition.
\begin{defn}
A \textbf{vertex 3-polytope} $P$ is equivalent to the following data.\\
\indent i) A set $S=\{1, ..., N\}$ of $N$ distinct boundaries (2-planes) in the above sense.  \\
\indent ii) $N$ cyclic lists $\{l_i=(j_{i1}, ..., j_{in_i})\}_{1\leq i\leq N}$ with each $j_{ik}\in S$, such that for all $i,k\in S$, $E_{i;k}=-E_{k;i}$.\\
\end{defn}
\vspace{-5mm}
\noindent For example, referring to the lists in (\ref{eq:3SimplexLists}), we have $E_
{2;1}=[3(12)4]=-[4(12)3]=-E_{1;2}.$  The rest can be checked explicitly, though we know the conditions will all be satisfied since the lists were derived from Figure \ref{fig:3Simplex} in such a way as to guarantee this fact.\\
\indent We note that vertex 3-polytopes can be just as ``disconnected'' and generic as vertex 2-polytopes can be.  Each two face of a 3-polytope can be an arbitrarily complicated 2-polytope, and in particular we have no notion of convexity since the boundaries of a 3-polytope can be disconnected.  This is one of the characteristics distinguishing this definition of polytope from that found in the mathematical literature.

\subsection{3D Volumes and the Vertex Calculus}
For the sake of brevity we carry our discussion of volumes of 3-polytopes over from the two-dimensional case strictly by analogy, though more detailed treatments do exist.  All proofs in this section are left out as they are similar in approach to the two-dimensional case.  We begin by defining the volume of a 3-simplex: 
\begin{equation} \label{eq:3SimplexVolume}
A_{3-simplex}=\frac{1}{6}\frac{\langle 1234\rangle^3}{\langle 123P_0\rangle\langle 234P_0\rangle\langle 341P_0\rangle\langle  412P_0\rangle}\equiv [1234].
\end{equation}
Analogously to (\ref{eq:TriangleArea}), we have defined $\langle ijkl\rangle \equiv \epsilon_{\alpha\beta\gamma\delta}Z_i^{\alpha}Z_j^{\beta}Z_k^{\gamma}Z_l^{\delta}.$   We note---as it will be important later---that if the four 2-planes defined by $Z_1^{\alpha},..., Z_4^{\alpha}$ intersect in a common point, then the $\{Z_i^{\alpha}\}_{1\leq i\leq 4}$ are linearly dependent and so $[1234]$ vanishes.

The volume $A(\{l_i\})$ of a general vertex 3-polytope defined by the lists $\{l_i=(j_{i1}, ..., j_{in_i})\}$ is defined to be
\begin{equation}
A(\{l_i\})\equiv\frac{2}{3!}\sum_{i=1}^N\sum_{k=1}^{n_i}[ij_{ik}j_{i(k+1)}B],
\label{eq:Arbitrary3PolytopeVolume}
\end{equation}
 which can be shown to be independent of the reference boundary $B^{\alpha}\in \mathbb{CP}^3$ and where the sum on $k$ (but not on $i$) is cyclic.  This is simply a particular triangulation of our polytope, where the prefactor $\frac{2}{3!}$ comes from the fact that, due to the constraints on the lists $\{l_i\}$, we are summing over each simplex once for every even permutation of $i,j_{ik},j_{i(k+1)}$ in (\ref{eq:Arbitrary3PolytopeVolume}).  We later use (\ref{eq:Arbitrary3PolytopeVolume}) to confirm that our vertex calculus obtains the correct volume, though as in the two-dimensional case we will see that this calculus is independent of any particular triangulation.\\
\indent We now introduce the three-dimensional analogues of the $V[\cdot\cdot][\cdot\cdot]$ objects.  For any six boundaries $1, ...,6$, we define
\begin{equation}
V[12][34][56]\equiv [1235]-[1236]-[1245]+[1246].
\end{equation}
Using the three-dimensional analogue of (\ref{eq:FundamentalIdentity}) given by \begin{equation}[1235]-[2345]+[3415]-[4125]=[1234], \end{equation} one can show that $V[\cdot\cdot][\cdot\cdot][\cdot\cdot]$ is fully antisymmetric both in its individual $[\cdot\cdot]$ entries as well as under swapping the $[\cdot\cdot]$'s themselves.  As shown in Appendix \ref{app:3Cube}, $V[12][34][56]$ corresponds to the volume of a cube, just as $V[12][34]$ corresponds to the volume of a quadrilateral in the two-dimensional case.  We define the brackets $\{...|...|...\}$ to be one-half times the (non-normalized) antisymmetrization of the labels that are excluded from the vertical bars.  For example, 
\begin{equation}
V[\{ij][k\}P][QR]\equiv V[ij][kP][QR]+V[jk][iP][QR]+V[ki][jP][QR]
\end{equation}
and 
\begin{align}
V[\{ij][k|P|][l\}Q]&\equiv V[\{ij][k\}P][lQ]-V[\{jk][l\}P][iQ]\\ \nonumber
&+V[\{kl][i\}P][jQ]-V[\{li][j\}P][kQ].
\end{align}
Then, given a set $S=\{1, ..., N\}$ of $N$ distinct boundaries and two fixed reference boundaries $P^{\alpha}$ and $Q^{\alpha}$, we define the following vertex objects for each $i,j,k\in S$:
\begin{equation}
F_{ijk}\equiv \sum_{l\neq i,j,k}^NV[\{ij][k\}P][lQ].
\end{equation}
Each $F_{ijk}$ is clearly antisymmetric in its subscripts, and they can be straightforwardly shown to individually be independent of $P^{\alpha}$ (though they are dependent on $Q^{\alpha}$, as in the two-dimensional case).  We then have the following cohomological identity:
\begin{equation}
\rho_{[i}F_{jkl]}\equiv F_{ijk}-F_{jkl}+F_{kli}-F_{lij}=3![ijkl]. \label{eq:main3DIdentity}
\end{equation}
We again see the resemblance to the $\check{\text{C}}$ech cocycle condition now for the second cohomology group as well as the many-fold covering of the simplex.\\

The following result defines the three-dimensional vertex calculus.  With $A(\{l_i\})$ being the volume of the vertex 3-polytope defined by the lists $\{l_i\}$ as in (\ref{eq:Arbitrary3PolytopeVolume}), one can show that
\begin{equation}
A(\{l_i\})=\frac{2}{(3!)^2}\sum_{i=1}^N\sum_{k=1}^{n_i}F_{ij_{ik}j_{i(k+1)}}\label{eq:3DCalculus},
\end{equation}
where the sum on $k$ (but not on $i$) is cyclic.  This sum of vertex objects is therefore independent of $Q^{\alpha},$ just as in the two-dimensional case.  The three-dimensional calculus is therefore to go along each cyclic list (one for each boundary plane) and add up the corresponding $F_{ijk}$ at each vertex.  In other words, we simply go through the two-dimensional calculus on each face with the $F_{ij}$ objects replaced by $F_{ijk}$ objects in the appropriate way.\\
\indent The proof of (\ref{eq:3DCalculus}) is similar to the two-dimensional case, though relies heavily on the constraints imposed upon the lists by their defining a 3-polytope, and thus does involve some added care.  For brevity, however, we leave this proof out since it is the \textbf{utility} of this formalism that we want to focus on.  Namely, we again have an expression of the volume in terms of objects that label the genuine vertices of the underlying polytope, and we obtain this expression without any reference to a triangulation.  We also have a cohomological identity that allows us to recover any triangulation in the same manner as in two dimensions.  Finally, we note that the same type of double sum is performed on the $F_{ijk}$'s in (\ref{eq:3DCalculus}) as is performed on the $[abcd]$ objects in (\ref{eq:Arbitrary3PolytopeVolume}).  This does not mean that the respective summands behave similarly, for proving the equivalence of these two sums is non-trivial, but it does mean that we gain all of the benefits of viewing the vertex objects as the atomic objects of the formalism without the cost of introducing more terms in the sums.  We leave an explicit three-dimensional example for Appendix \ref{app:3Cube}.\\

\section{Vertex 4-Polytopes}
We now seek to define 4-polytopes in terms of cyclic lists, as we would then expect a useful vertex calculus to be obtained thereafter.  The key observation in this regard is to view 4-polytopes as a set of 3-polytopes---one for each hyper-face---``glued-together'' along their two-dimensional faces in such a way as to make the ``boundary of the boundary'' of the polytope vanish.\\
\indent As usual, we suppose we have a set $S=\{1, ..., N\}$ of $N$ distinct elements $Z_1^{\alpha}, ..., Z_N^{\alpha}\in \mathbb{CP}^4$, defining $N$ distinct 3-planes (i.e., linearly embedded $\mathbb{CP}^3$'s) in the dual space, and we refer to $S$ as the set of boundaries.  The intersection of any two distinct 3-planes determines a 2-plane (i.e., a linearly embedded $\mathbb{CP}^2$), the intersection of any three distinct 3-planes determines a (complex projective) line denoted by $i\cap j\cap k$, and the intersection of any four distinct 3-planes determines a point, or vertex, denoted by $\{i,j,k,l\}.$  An oriented edge is now of the form $[i(jkl)m]$ and denotes the abstract instructions $\{i,j,k,l\}\rightarrow \{j,k,l,m\}$, meaning to go from the vertex $\{i,j,k,l\}$ to the vertex $\{j,k,l,m\}$ along the line defined by the intersection of the three 3-planes labelled by the three labels common to the two vertices.\\
\indent We now say that a vertex 4-polytope $P$ is determined by $N^2$ cyclic lists $\{l_{ij}=(k_{ij1}, ..., k_{ijn_{ij}})\}$ where each $k_{ijl}\in S$ and $n_{ij}$ is the length of the list $l_{ij}$.  We view the list $l_{ij}$ as the list defining the 2-polytope obtained by first restricting the 4-polytope $P$ to the three-dimensional polytope $P_i$ sitting on the $i^{th}$ face, and then restricting $P_i$ to the $j^{th}$ face to get the 2-polytope $P_{ij}$. The lists $l_{ii}$ labelled by the same boundary are by convention empty.\\
\indent We define the edge sets $E_{ij}\equiv E(l_{ij})$ in the usual way:
\begin{equation}
E_{ij}=\sum_{l=1}^{n_{ij}}[k_{ij(k-1)}(ijk_{ijl})k_{ij(k+1)}],
\end{equation}
with the sum on $l$ cyclic, so that $E_{ij}$ corresponds to the edge set of the 2-polytope $P_{ij}$.  We then have that the 3-polytope $P_i$ is defined by the $(N-1)$ lists $\{l_{ij}\}$ with $j\in S-\{i\}$, and accordingly by the edge sets $E_{ij}$ with $j\in S-\{i\}$.  Namely, we simply fix the first subscript.  Requiring that $P_i$ is indeed a 3-polytope for each $i\in S$, we simply carry over the three-dimensional constraint which is that for all $i,j,k\in S$, $E_{ij;k}=-E_{ik;j}$.  In order to view these 3-polytopes as being properly ``glued'' along their boundary 2-polytopes, we note that if we first restrict $P$ to $P_i$ and then to $P_{ij}$, we are looking at the mirror image of the 2-polytope that we would get by first restricting to $P_j$ and then to $P_{ji}$.  Thus, we need to impose that for all $i,j\in S$, $E_{ij}=-E_{ji}$.  Combining these two constraints on the edge sets $\{E_{ij}\}$, we are motivated to make the following definition of vertex 4-polytopes.
\begin{defn}
A \textbf{vertex 4-polytope} $P$ is equivalent to the following data.\\
\indent i) A set $S=\{1, ..., N\}$ of $N$ distinct boundaries (3-planes) in the above sense.  \\
\indent ii) $N\choose 2$ cyclic lists $\{l_{ij}=(k_{ij1}, ..., k_{ijn_{ij}})\}_{1\leq i,j\leq N}$ with each $k_{ijl}\in S$, such that for all $i,j,l\in S$, $E_{ij;l}=(-1)^{|\sigma|}E_{\sigma(ij;l)}$ where $\sigma\in S_3$ is any permutation of 3 objects.\\
\end{defn} 
\vspace{-4mm}
\noindent At this point it is now clear how to extend our definition of vertex polytopes to any dimension, and we do so explicitly in Appendix \ref{app:DPolytopes}.

\subsection{4D Volumes and the Vertex Calculus}
In what follows we suppose the cyclic lists $\{l_{ij}=(k_{ij1}, ..., k_{ijn_{ij}})\}$ with $k_{ijl}\in S=\{1, ... N\}$ define a vertex 4-polytope $P$. We denote by $[12345]$ the volume of the 4-simplex bounded by the boundaries $Z_1^{\alpha}, ..., Z_5^{\alpha}\in \mathbb{CP}^4$, which is given as the four-dimensional analogue of (\ref{eq:TriangleArea}) and (\ref{eq:3SimplexVolume}) (see Ref.~\cite{NoteOnPolytopes} for more details).  We then define the volume $A(\{l_{ij}\})$ of $P$, an arbitrary vertex 4-polytope, to be
\begin{equation} \label{eq:Arbitrary4DPolytopeVolume}
A(\{l_{ij}\})\equiv \frac{2}{4!}\sum_{i=1}^N\sum_{j=1}^N\sum_{l=1}^{n_{ij}}[ijk_{ijl}k_{ij(l+1)}B]
\end{equation}
for some reference boundary $B^{\alpha}\in \mathbb{CP}^4$ (though it can be shown that $A(\{l_{ij}\})$ is independent of our choice of $B^{\alpha}$) and where the sum on $l$ (but not on $i$ or $j$) is cyclic.  We then define the four-dimensional analogue of the $V[\cdot\cdot][\cdot\cdot]$ objects, where for any eight boundaries $1, ..., 8$ we have
\begin{align}
V[12][34][56][78]\equiv & [12357]-[12358]-[12367]+[12368] \\ \nonumber
&-[12457]+[12458]+[12467]-[12468],
\end{align}
and we define our vertex objects as follows:
\begin{equation}
F_{ijkl}=\sum_{m\neq i,j,k,l}^{N}V[\{ij][k|P_1|][l\}P_2][mQ]
\end{equation}
for some reference boundaries $P_1^{\alpha}, P_2^{\alpha}$, and $Q^{\alpha}$.  It can be shown that each $F_{ijkl}$ is individually independent of $P_1^{\alpha}$ and $P_2^{\alpha}$, but is dependent on $Q^{\alpha}$.  It can also be shown, though, that the sum
\begin{equation}
\sum_{i=1}^N\sum_{j=1}^N\sum_{l=1}^{n_{ij}}F_{ijk_{ijl}k_{ij(l+1)}},
\end{equation}
with the sum on $l$ (but not on $i$ or $j$) cyclic in the usual sense, is independent of our choice of $Q^{\alpha}$.  Indeed, our four-dimensional calculus carries through just as it does in the lower dimensions with the following result, with $A(\{l_{ij}\})$ as in (\ref{eq:Arbitrary4DPolytopeVolume}):
\begin{equation}\label{eq:main4DResult}
A(\{l_{ij}\})=\frac{2}{(4!)^2}\sum_{i=1}^N\sum_{j=1}^N\sum_{l=1}^{n_{ij}}F_{ijk_{ijl}k_{ij(l+1)}}.
\end{equation}
The proof of this result relies heavily on the constraints placed on the cyclic lists in order for them to form a genuine 4-polytope, but instead of giving the proof we will instead focus on the utility of this result.  We first note that the right hand side of (\ref{eq:main4DResult}) is manifestly dependent only on the vertices of the underlying polytope, via the subscripts of each vertex object, just as in the two- and three-dimensional cases.  Moreover, using the following cohomological identity (whose proof we omit):
\begin{equation}\label{eq:key4DIdentity}
\rho_{[i}F_{jklm]}\equiv F_{ijkl}+F_{jklm}+F_{klmi}+F_{lmij}+F_{mijk}=4![ijklm],
\end{equation}
we can obtain \textbf{any} triangulation we desire by using (\ref{eq:key4DIdentity}) along with (\ref{eq:main4DResult}), just as we could in dimensions two and three.

We note that equation (\ref{eq:key4DIdentity}) is similar to equation (10) of Ref.~\cite{NoteOnPolytopes}, given by \begin{equation}\partial [ijklm]=[ijkl]+[jklm]+[klmi]+[lmij]+[mlij], \label{eq:NimaEquation}\end{equation} describing the boundary of a simplex and encoding where the poles of $[ijklm]$ are.  We note, though, that while (\ref{eq:key4DIdentity}) encodes the poles of $[ijklm]$ as well, it is also a genuine equality between the vertex objects and (a multiple of) the volume of a simplex.  Therefore, the objects on the left of equation (\ref{eq:key4DIdentity}) are fundamentally different than those on the right of (\ref{eq:NimaEquation}).
\subsubsection{Lists from Triangulations}
We have seen how we can express our volumes in a triangulation-independent and manifestly vertex-dependent way once we know the cyclic lists.  However, in practice (using BCFW recursion, for example) we start with a particular triangulation and not a tabulation of the cyclic lists defining a polytope.  It is therefore worthwhile to see how we can extract the lists from any particular triangulation.  By doing this, we find that once we are given any particular triangulation, we can recover all of the information about the polytope---as well as any of its lower dimensional boundaries---using this calculus.  This is best seen via an example.\\
\indent We consider a 4-simplex with the obvious triangulation being simply $[12345]$.  Via (\ref{eq:key4DIdentity}), we see that
\begin{equation}\label{eq:4DSimplexListFromTriangle}
A(\{l_{ij}\})=[12345]=\frac{1}{4!}(F_{1234}+F_{2345}+F_{3451}+F_{4512}+F_{5123}),
\end{equation}
where we do not yet know the lists $\{l_{ij}\}$ defining the 4-simplex.  We do know from (\ref{eq:main4DResult}), however, that (\ref{eq:4DSimplexListFromTriangle}) must be a sum over cyclic lists.  Therefore, for example, by writing all of the vertex objects in (\ref{eq:4DSimplexListFromTriangle}) with 1 as the left-most subscript and 2 as the second-to-left-most subscript (while keeping track of relative minus signs via the total antisymmetry of the $F_{...}$'s), we immediately read off that $l_{12}=(3,4,5)$.  The other lists can be read off similarly. \\
\indent  If one is solely interested in the volume of a particular simplex, there is no need for obtaining all of the cyclic lists or for expanding the volume out as a sum of vertex objects---one would simply write $[abcde]$.  However, the process of reading off cyclic lists from a particular triangulation allows for the full utility of the vertex calculus for arbitrarily complex polytopes in a straightforward way.

\subsubsection{Going to the Boundary}
One of the benefits of using the vertex formalism in dimensions higher than two is that we can readily obtain the information (i.e., the cyclic lists and the volumes) of any of the lower dimensional boundary polytopes.  For example, suppose we are given a vertex 4-polytope $P$ defined by the lists $\{l_{ij}\}$ with entries in $S=\{1, ...,N\}$. The boundaries $\{Z_{i}^{\alpha}\}$ are all elements of $\mathbb{CP}^4$ and therefore have five homogeneous coordinates.  Accordingly, we define the volume of a 3-simplex defined by the boundaries $i,j,k,l$ restricted to the boundary $I\in S$  as follows:
\begin{equation}
[ijkl]_I\equiv \frac{1}{6}\frac{\langle ijklI\rangle^3}{\langle ijkIP_0\rangle \langle jklIP_0\rangle \langle kliIP_0\rangle \langle lijIP_0\rangle}.
\end{equation}
The objects $[ijkl]_I$ satisfy all of the same algebraic properties as the usual three-dimensional $[ijkl],$ so our three-dimensional calculus carries directly through by defining the new objects $\{F^I_{ijk}\}$ where every $[ijkl]$ is simply replaced by $[ijkl]_I$.  Then, if we want to calculate the volume of the 3-polytope that is the restriction of $P$ to the $I^{th}$ face, we simply apply the three-dimensional calculus to the lists $\{l_{Ij}\}$ with $j\in S-\{I\}$ using the $F^I_{ijk}$ objects. The same can be said about obtaining the area of any boundary 2-polytope by defining the area of a 2-simplex defined by the boundaries $i,j,k$ and \textbf{restricted to} the $I^{th}$ and then to the $J^{th}$ face of $P$ as
\begin{equation}
[ijk]_{IJ}\equiv \frac{1}{2}\frac{\langle ijkJI\rangle^2}{\langle ijJIP_0\rangle \langle jkJIP_0\rangle \langle kiJIP_0\rangle}.
\end{equation} 
The $[ijk]_{IJ}$ objects satisfy all of the algebraic properties that the two-dimensional $[ijk]$ objects satisfy, so by defining $F^{IJ}_{ij}$ in the obvious way and applying the two-dimensional calculus to the list $l_{IJ}$, we get the area of the polygon $P_{IJ},$ which is the restriction of $P$ to the $I^{th}$ and then to the $J^{th}$ face.

\subsection{Applications to $M_n^{NMHV}$}
We can use the vertex calculus and (\ref{eq:key4DIdentity}) to obtain a new, manifestly local, and triangulation-independent expression for $M_n^{NMHV},$ the $n$-point NMHV tree amplitude in planar $\mathcal{N}=4$ sYM theory discussed at the beginning of this note.  Simply for notational convenience we rewrite $Z_*^{\alpha}$ as $Z_0^{\alpha}$ in (\ref{NMHVTree}), and we find
 \begin{align}
 M_{n}^{NMHV}&=\frac{1}{2}\sum_{i,j=1}^n[0i(i+1)j(j+1)]\\
 &=\frac{1}{2\cdot 4!}\sum_{i,j=1}^n(F_{0i(i+1)j}+F_{i(i+1)j(j+1)}+F_{(i+1)j(j+1)0}+F_{j(j+1)0i}
 +F_{(j+1)0i(i+1)})\\ \label{eq:NMHVBetterForm}
 &=\frac{1}{2\cdot 4!}\sum_{i,j=1}^nF_{i(i+1)j(j+1)}.
 \end{align}
In the third equality any term with a $0$ subscript cancels after relabeling of the $i$'s and $j$'s, which we are permitted to do due to the cyclicity of the sum.  Thus the dependence on the reference boundary $0$ manifestly drops out, while the fact that the underlying polytope only has vertices of the form $\{i,i+1,j,j+1\}$---which is the statement that this polytope represents a \emph{local} amplitude---remains manifest due to the fact that these are the only subscripts of the remaining vertex objects.  We can use (\ref{eq:key4DIdentity}) to obtain any valid triangulation we desire from (\ref{eq:NMHVBetterForm}).  

We emphasize that it is the vertex objects that can be used to uniquely express these amplitudes.  For example, for $n=6$ we can write 
\begin{equation}
M_6^{NMHV}=[12345]+[12356]+[13456],
\end{equation}
as well as
\begin{equation}
M_6^{NMHV}=[23456]+[23461]+[24561],
\end{equation}
with these seemingly different expressions arising from two different BCFW shifts, or by plugging in $Z_1^{\alpha}$ and $Z_2^{\alpha}$ (respectively) for $Z_*^{\alpha}$ in (\ref{NMHVTree}).  The equality of these two triangulations is made manifest by using (\ref{eq:key4DIdentity}) on each of them to find that, in both cases, one obtains
\begin{align}
M_{6}^{NMHV}=&\frac{1}{4!}(F_{1234}+F_{2345}+F_{2356}+F_{5612}+F_{6123} \label{uniqueSixPoint}\\ \nonumber
&+F_{3456}+F_{4561}+F_{6134}+F_{1245}).\\ \nonumber
\end{align}
Indeed, one would arrive at (\ref{uniqueSixPoint}) by using (\ref{eq:key4DIdentity}) on any valid triangulation of $M_{6}^{NMHV}.$  Therefore, since the volume of any polytope is uniquely expressed in terms of vertex objects, these objects manifest non-trivial identities amongst triangulations.  These identities are obtained in the Grassmannian picture of Refs.~\cite{BigGrassmannian, MasonSkinner, DualityForSMatrix} via the use of GRTs.  In the vertex calculus, however, these identities are manifested simply by the fact that when equivalent triangulations are expressed in terms of vertex objects, the resulting expressions are identical.  This is also done while encoding all of the relevant information about the underlying polytope via the subscripts of these vertex objects.

To see this, let us use this formalism to also find the volume of the 3-polytope $P_2$ sitting on the boundary defined by $Z_2^{\alpha}$ of $M_{6}^{NMHV},$ as was done in Ref.~\cite{SpuriousPoles}.  We can read off the lists for this 3-polytope from (\ref{uniqueSixPoint}) to find the following:
\begin{align}
l_{21}&=(4,3,6,5)\\ \nonumber
l_{23}&=(1,4,5,6)\\ \nonumber
l_{24}&=(3,1,5)\\ \nonumber
l_{25}&=(4,1,6,3)\\ \nonumber
l_{26}&=(5,1,3).
\end{align}
We then use the three-dimensional calculus on these five lists with the $F^2_{ijk}$ vertex objects, where the superscript 2 represents the restriction to this boundary hyperplane.  Denoting the volume of $P_2$ by $A_2(\{l_{2i}\})$, we find
\begin{align}\nonumber
A_2(\{l_{2i}\})=&\frac{1}{18}\sum_{i=1}^6\sum_{l=1}^{n_{2i}}F^2_{il(l+1)}\\ 
=&\frac{1}{6}(F^2_{143}+F^2_{136}+F^2_{165}+F^2_{154}+F^2_{345}+F^2_{356})\label{eq:CyclicPolytopeFace}.
\end{align}
We now use (\ref{eq:main3DIdentity}) to find
\begin{align}
A_2(\{l_{2i}\})=&[1365]_2-[1345]_2,
\end{align}
which is in line with the results of Ref.~\cite{SpuriousPoles}, and is just one possible triangulation that can be obtained from (\ref{eq:CyclicPolytopeFace}).
\subsubsection{Triangulations Without Spurious Vertices}

We can obtain infinitely many triangulations from (\ref{eq:CyclicPolytopeFace}) by allowing ourselves to introduce spurious boundaries as well as spurious vertices, by for example adding and subtracting $F^2_{127}$ where the boundary 7 is some new reference boundary and then simplifying using (\ref{eq:key4DIdentity}).  However, we can also be more careful and introduce spurious boundaries in such a way that we can triangulate our space without introducing spurious vertices.  This was done in Ref.~\cite{NoteOnPolytopes} for the boundary 3-polytope $P_2$ (both for the $n=6$ case that we just considered as well as the general $n$ case) and in four dimensions for the full $M_n^{NMHV}$ polytope.  The procedure in Ref.~\cite{NoteOnPolytopes} relies heavily on geometric insight to ``chop up'' the various polytopes into simplices without introducing new vertices.  Here, we focus solely on the $n=6$ boundary 3-polytope $P_2$ and recover the particular spurious-vertex-free triangulation obtained in Ref.~\cite{NoteOnPolytopes} in a purely algebraic fashion.

\begin{figure}[t!]
\centering
\captionsetup{width=0.8\textwidth}
    \includegraphics[scale=0.3]{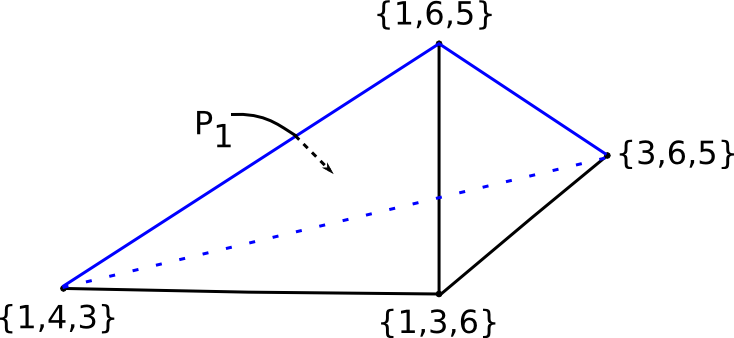}
        \caption{A depiction of the newly defined plane when triangulating the three-polytope $P_2$ without introducing spurious vertices.}\label{fig:TriangulationWOVertices}
\end{figure}

The general procedure is schematically as follows.  We begin with a sum of vertex objects denoting the physical vertices of the underlying polytope.  Suppose there are $m$ such vertices.  We pick one physical vertex arbitrarily---corresponding to a particular vertex object---and ``triangulate it away'' by choosing three (in three dimensions) other vertex objects whose subscripts share precisely two of the three labels of the original vertex object.  This corresponds to picking a total of four non-coplanar vertices, with the initially chosen vertex being connected by edges of the underlying polytope to the other three vertices.   We then introduce the plane through these latter three vertices into our list of boundaries, add the volume of the simplex defined by this (possibly new) plane and the four vertices we have chosen, and write out the remaining $F_{...}$ terms.  As we will see by example, this gives the original volume of the polytope with $m$ physical vertices as the sum of a simplex (with only physical vertices) and a polytope with $m-1$ physical vertices, in general depending on the plane just introduced. The triangulation that one is left with after this procedure is not unique as there are many arbitrary choices made along the way (as in, for example, which vertex to ``triangulate away'' at any given step), but it will be guaranteed to have no spurious vertices.  Let us see how this works via an example.
\\
\indent Focusing on the boundary 3-polytope $P_2$ of $M_6^{NMHV}$, we suppress the $2$ superscript on the vertex objects and simply note that all of the following takes place within the three-dimensional restriction to the boundary plane defined by $Z_2^{\alpha}$.  From (\ref{eq:CyclicPolytopeFace}) we can arbitrarily choose to first triangulate away the vertex $\{1,3,6\}$.  We also see that $\{1,3,4\}, \{1,5,6\},$ and $\{3,5,6\}$ are all physical vertices sharing precisely two labels with $F_{136}$. Accordingly, we define the plane $P_1^{\alpha}$ to be the plane through these three vertices:
\begin{equation}
P_1^{\alpha}\equiv \ \rm{plane}\ \rm{through}\ \{1,3,4\}, \{1,5,6\}, \{3,5,6\}.
\end{equation}
We then see that certain vertices are labelled in different equivalent ways, as for example $\{1,3,4\}=\{P_1,1,3\}=\{P_1,3,4\}=\{P_1,1,4\}$.  From the discussion immediately following (\ref{eq:3SimplexVolume}) and the definition of $P_1^{\alpha}$, we find
\begin{align}
[134P_1]=[156P_1]=[356P_1]=0, 
\end{align}
which then gives, via (\ref{eq:main3DIdentity}), the following identities:
\begin{align} \label{eq:TriangulationWOVertices1}
F_{143}&=F_{P_114}+F_{P_143}+F_{P_131},\\ \nonumber F_{165}&=F_{P_116}+F_{P_165}+F_{P_151},\\ \nonumber
F_{563}&=F_{P_156}+F_{P_163}+F_{P_135},
\end{align}
where in each line all three terms on the right hand side label the same vertex as the term on the left hand side.  We then also have
\begin{equation}
F_{136}=3![136P_1]+F_{P_113}+F_{P_136}+F_{P_161}, \label{eq:TriangulationWOVertices2}
\end{equation}
and the vertices of $[136P_1]$ are all physical.  We depict the geometry behind this algebra in Figure \ref{fig:TriangulationWOVertices}.

What we have done is use the vertex information given to us from (\ref{eq:CyclicPolytopeFace}) to eliminate the vertex $\{1,3,6\}$ by defining the plane through the end points of three edges connecting $\{1,3,6\}$ to other physical vertices (all of which being known from (\ref{eq:CyclicPolytopeFace})).  It is then no surprise that all of the vertices of $[136P_1]$ are physical.

By plugging (\ref{eq:TriangulationWOVertices1}) and (\ref{eq:TriangulationWOVertices2}) into (\ref{eq:CyclicPolytopeFace}), making the immediate cancellations, and recalling that we have suppressed the 2 superscript, we find
\begin{equation}
A_2(\{l_{2i}\})= [136P_1]+\frac{1}{6}((F_{P_114}+F_{P_143})+(F_{P_165}+F_{P_151})+(F_{P_156}+F_{P_135})
+F_{154}+F_{345})
\end{equation}
where we have lumped two vertex objects together if they label the same (physical) vertex.  Thus we see that we have now expressed the volume as the sum of a simplex with all physical vertices and a new polytope with only five physical vertices.  The general idea, then, is to repeat this process until the ``remainder'' polytope has only four vertices and then employ the identities amongst the newly defined planes $P_i^{\alpha}$ to express this remainder polytope as the volume of a simplex (which is guaranteed to have only physical vertices).  For completeness, we finish our current example.\\
\indent We now define $P_2^{\alpha}$ as 
\begin{equation}
P_2^{\alpha}\equiv  \ \text{plane}\ \text{through}\ \{P_1,1,5\}=\{1,5,6\}, \{P_1,1,4\}=\{1,3,4\},\{3,4,5\}.
\end{equation}
Then by reading off the identities analogous to (\ref{eq:TriangulationWOVertices1}) for $F_{1P_15}, F_{14P_1},$ and $F_{345}$, and the result analogous to (\ref{eq:TriangulationWOVertices2}) for $F_{154}$ (``triangulating away'' this vertex), putting it all together and employing one final use of (\ref{eq:main3DIdentity}), we find
\begin{equation}\label{eq:TriangulationWOVertices3}
A_2(\{l_{2i}\})=[136P_1]_2+[154P_1]_2+[35P_1P_2]_2,
\end{equation}
where we have reinstated the subscript 2.  The vertices of each simplex in (\ref{eq:TriangulationWOVertices3}) can be readily checked to all be physical (i.e., non-spurious) vertices.  The expression (\ref{eq:TriangulationWOVertices3}) agrees with what was found in Ref.~\cite{NoteOnPolytopes}.  In our formalism this triangulation (as well as any other) can be obtained from (\ref{eq:CyclicPolytopeFace}), or equivalently from any particular triangulation (by first reconstructing the cyclic lists), in a purely algebraic fashion. This example and the statement of the general procedure makes us believe that a general algorithm (for any polytope in any dimension) for moving from a sum of vertex objects to a sum of volumes of simplices with no spurious vertices should exist.  This would be useful both for higher dimensional polytopes as well as more complicated polytopes.

\section{Summary and Outlook}
In this note we have developed a formalism in which we define polytopes via their boundary hyper-planes and cyclic lists describing their boundary two-dimensional faces. These polytopes are completely general---there is no restriction to connectedness or convexity---and are combinatorial in nature.  Additionally, we found that by defining certain sums of simplices $F_{ij}, F_{ijk}, F_{ijkl}, ...$, we can express the volumes of vertex polytopes in a manner that depends only on the vertices of the underlying polytope and is therefore independent of any choice of triangulation.  In fact, there is a unique expression for the volume of a polytope in terms of these vertex objects, which in turn manifests identities amongst various triangulations.  From this unique expression, any valid triangulation can be obtained algebraically using the ``cohomological'' identities  (\ref{eq:main2DIdentity}), (\ref{eq:main3DIdentity}), and (\ref{eq:key4DIdentity}).\\
\indent We also saw that by considering certain canonical subcollections of cyclic lists and defining canonical ``restricted'' vertex objects with respect to certain boundaries, we can immediately calculate the volume of any lower dimensional boundary polytope.  Finally, we saw that it is straightforward to use \textbf{any} particular triangulation of a polytope to obtain the cyclic lists for the underlying polytope.  Therefore, given any triangulation of a vertex $d$-polytope, one can algorithmically obtain the volumes of any lower dimensional boundary polytopes by first recovering the lists, restricting to the relevant lists, and using the restricted vertex calculus with these lists.\\
\indent These considerations all took place within $\mathbb{CP}^n$, which can be viewed as the simplest of all Grassmannia $G(k,n)$.  A relevant extension of these ideas would be to Grassmannia with $k>1$.  We believe that in order to make contact with these more complex spaces we must generalize the cohomological descriptions of the vertex objects.  The details of the cohomological considerations underlying the definition of the vertex objects will be left for a future note.  In short, though, in two dimensions these objects come from simple contour integrals of certain $\check{\text{C}}$ech cohomology classes of $\mathbb{CP}^2$.  Understanding this cohomological structure as well as its interaction with the base space (where the polytope lives) will likely be the key to understanding how this formalism generalizes to $G(k,n)$ and the ideas in Refs.~\cite{Amplituhedron,BigGrassmannian}. 

\section*{Acknowledgements}
I would like to thank Andrew Hodges, as well as Sir Roger Penrose, Lionel Mason, Zvi Bern, and Stephen Huggett for their suggestions.  I would like to thank Oxford University and the Mani L. Bhaumik Institute for Theoretical Physics for the resources they provided.  Finally, I would like to thank the Euretta J. Kellett Fellowship for providing my financial support during this research.

\appendix
\section{Example of Vertex Calculus} \label{app:2DCalculus}
\begin{figure}[t!]
\centering
\captionsetup{width=0.8\textwidth}
    \includegraphics[scale=0.3]{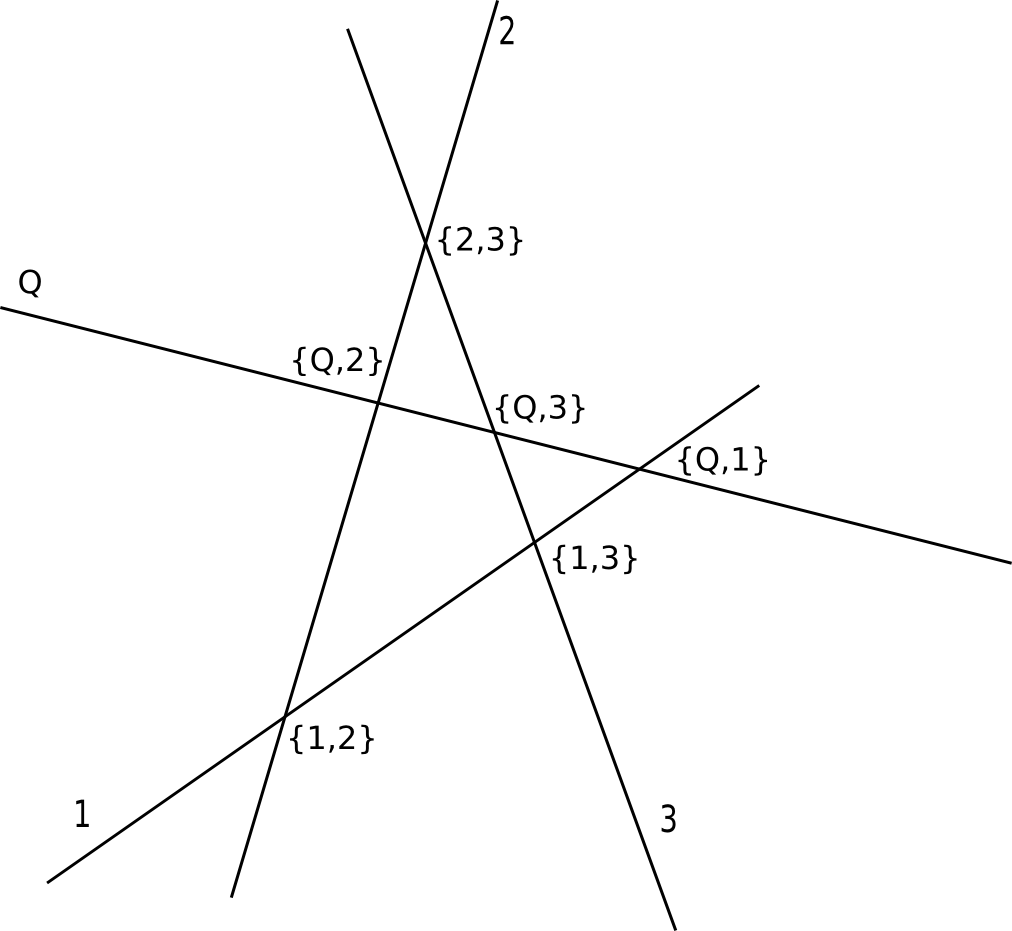}
        \caption{A collection of four boundaries, $\{1, 2, 3, Q\}.$}\label{fig:2Dexmp1}
\end{figure}

We consider the triangle, defined by the list $l=(1,2,3)$, so that the set of boundaries is $S=\{1,2,3\}$.  To define our $F_{ij}$'s, we must introduce a reference boundary $Q^{\alpha}$.  In Figure \ref{fig:2Dexmp1} we give a possible picture of these choices.

\begin{figure}[t!]
\centering
\captionsetup{width=0.8\textwidth}
    \includegraphics[scale=0.3]{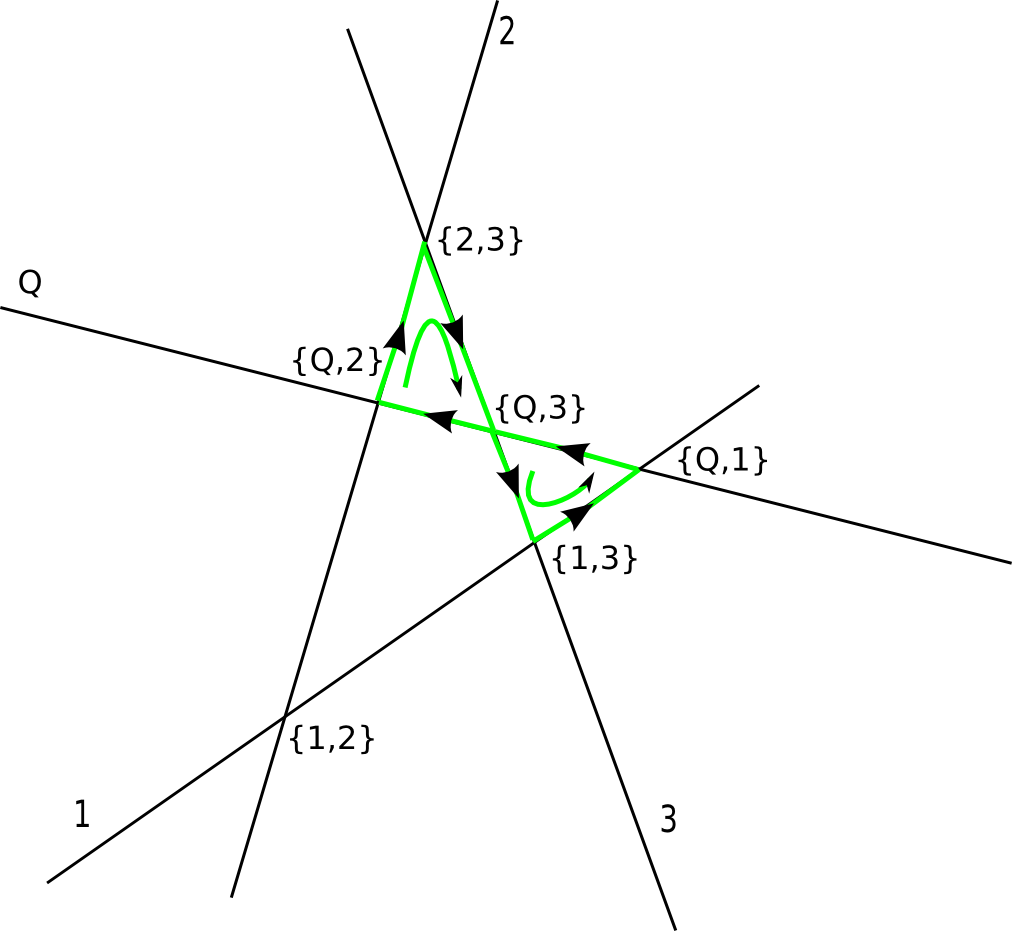}
        \caption{The area defined by the vertex object $F_{12}.$}\label{fig:2Dexmp2}
\end{figure}

\begin{figure}[t!]
\centering
\captionsetup{width=0.8\textwidth}
    \includegraphics[scale=0.3]{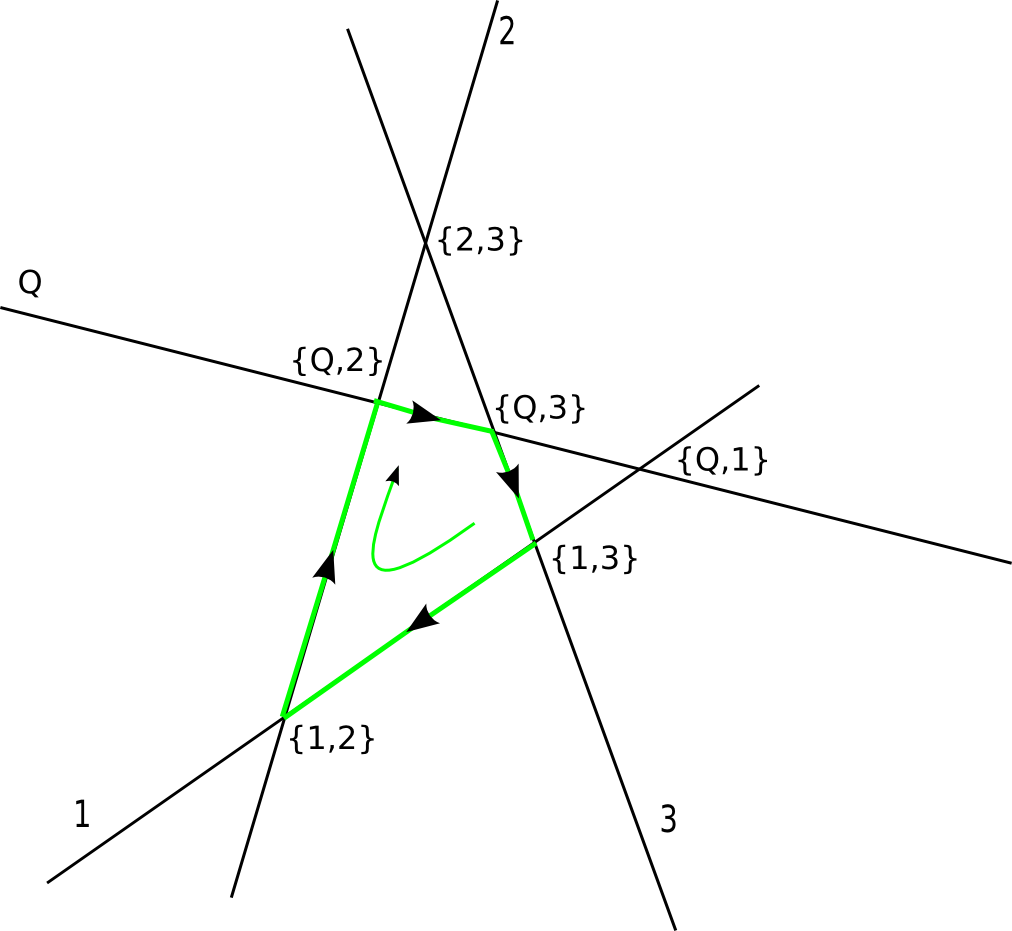}
            \caption{The area defined by the vertex object $F_{23}.$}\label{fig:2Dexmp3}
\end{figure}

\begin{figure}[t!]
\centering
\captionsetup{width=0.8\textwidth}
    \includegraphics[scale=0.3]{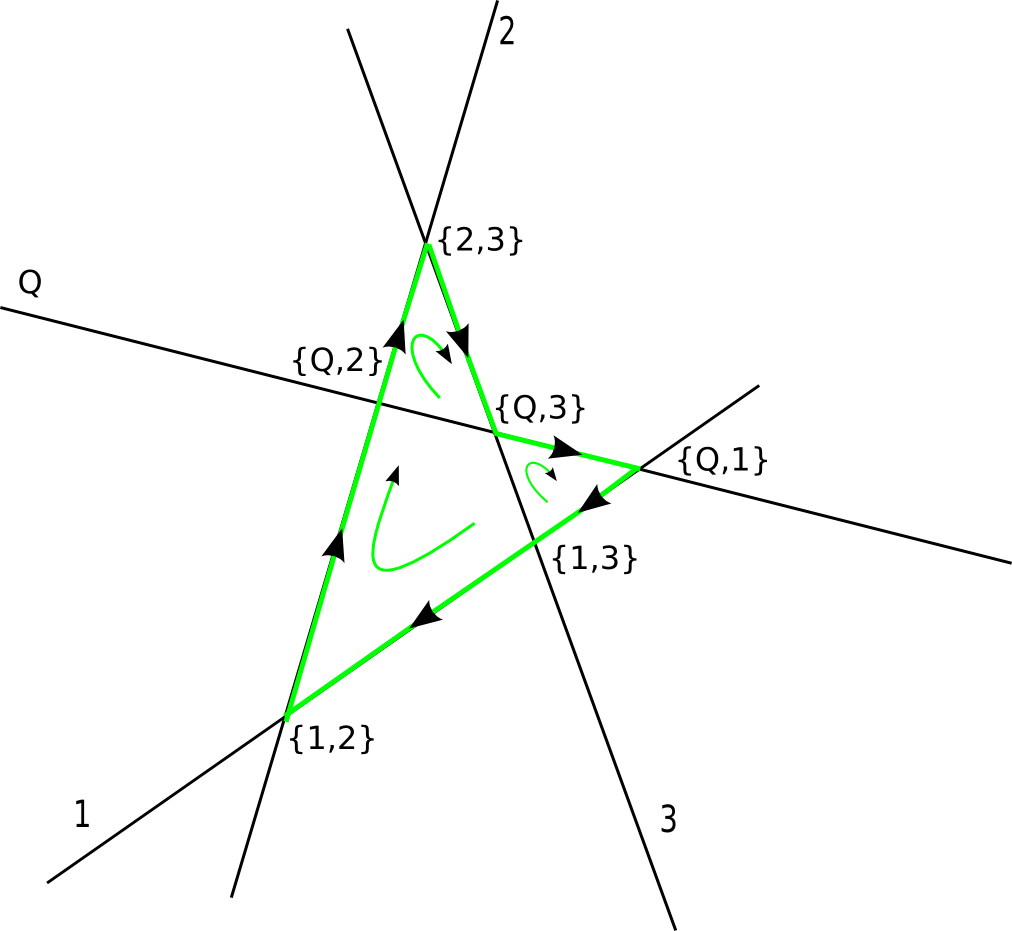}
            \caption{The area defined by the vertex object $F_{31}.$}\label{fig:2Dexmp4}
\end{figure}

With $S$ as above, i.e., not adding in another reference boundary $B^{\alpha}$ as in Theorem \ref{thm:Main2DThm}, we have 
\begin{align}
F_{12}=V[12][3Q]\\
F_{23}=V[23][1Q]\\
F_{31}=V[31][2Q].
\end{align}
Theorem \ref{thm:Main2DThm} says that we should consider the sum
\begin{equation}
\frac{1}{2}(F_{12}+F_{23}+F_{31})
\end{equation}
to get the area of the polygon defined by the list $l=(1,2,3)$.  Let us see what each of these terms corresponds to.  From the discussion following (\ref{eq:2DQuadrilateral}), we have that $F_{12}$ is the area of the polygon defined by the list $(1,Q,2,3)$.  This polygon is depicted in Figure \ref{fig:2Dexmp2}.
We now see that $F_{23}$ is the area of the polygon defined by the list $(2,Q,3,1)$, which is depicted in Figure \ref{fig:2Dexmp3}.  Finally, we see that $F_{31}$ is the area of the polygon defined by the list $(3,Q,1,2)$, depicted in Figure \ref{fig:2Dexmp4}.

By ``superposing'' the three polygons depicted in Figures \ref{fig:2Dexmp2}-\ref{fig:2Dexmp4} and cancelling the areas that have opposite orientations, we see that we are left precisely with 2 times the area of the triangle defined by the list $(123)$, and the dependence on $Q^{\alpha}$ drops out.  Had we included some other reference boundary $B^{\alpha}$ in $S$ to define $S'=S\cup \{B^{\alpha}\}$, then we would get new $F_{ij}$'s.  Namely, we would have
\begin{align}
F_{12}=V[12][3Q]+V[12][BQ]\\
F_{23}=V[23][1Q]+V[23][BQ]\\
F_{31}=V[31][2Q]+V[31][BQ],
\end{align}
and the analogues of Figures \ref{fig:2Dexmp2}-\ref{fig:2Dexmp4} would be correspondingly more complicated.  However, we can see immediately that the sum 
\begin{equation}
F_{12}+F_{23}+F_{31}
\end{equation}
would be left unaffected, since
\begin{equation}
V[12][BQ]+V[23][BQ]+V[31][BQ]=0
\end{equation}
using (\ref{eq:2DUsefulRelation}).  This is precisely the vanishing term in the second equality in the proof of Proposition \ref{prop:main2DProp}, and this is why we can add (but not subtract) as many boundaries to $S$ as we would like, or as is convenient, without having to worry about the sum of $F_{ij}$'s that we are interested in being affected.  Thus, we add $B^{\alpha}$ to the set of boundaries $S$ in Theorem \ref{thm:Main2DThm} because considering the objects $F_{iB}$ makes the proof simpler, but in practice we can deal only with $S$ and define our $F_{ij}$'s with respect to it.  We note that in this example the object $F_{1B}$ (as well as $F_{2B}$ and $F_{3B}$) can be defined using $S'$, and we would have $F_{1B}=V[1B][2Q]+V[1B][3Q]$ (and similarly for $F_{2B}$ and $F_{3B}$), but since $B$ does not make an appearance in the list $l=(1,2,3)$ under consideration, we can disregard these objects.

\section{Vertex 3-Cube} \label{app:3Cube}
\begin{figure}[t!]
\centering
\captionsetup{width=0.8\textwidth}
    \includegraphics[scale=0.3]{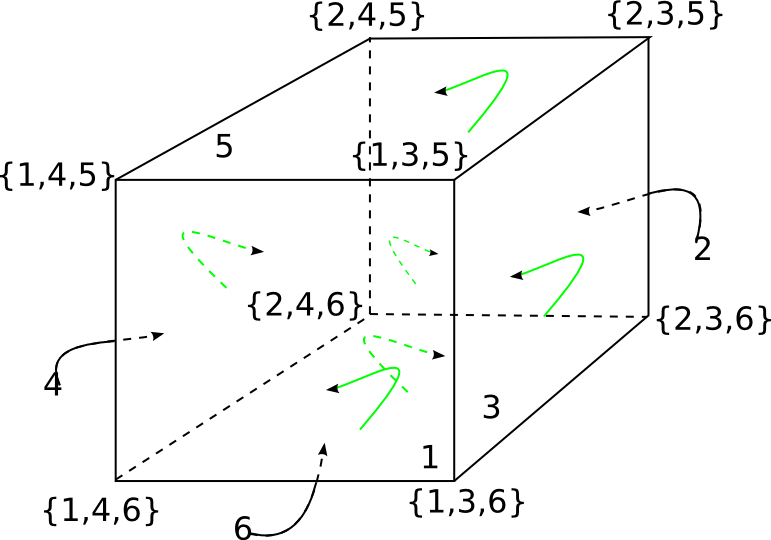}
        \caption{A three-dimensional cube with the green arrows depicting the cyclic list associated with its respective face.}\label{fig:Cube}
\end{figure}

We consider a ``cube,'' depicted in Figure \ref{fig:Cube} and defined by six boundaries, so that $S=\{1, 2, 3, 4, 5, 6\}$. 
 We read off from the picture that the following six lists (one for each 2-face) are the lists that reflect the (oriented) combinatorial properties of the cube:
\begin{align}
l_1&=(4,6,3,5)\\ \nonumber
l_2&=(5,3,6,4)\\ \nonumber
l_3&=(1,6,2,5)\\ \nonumber
l_4&=(5,2,6,1)\\ \nonumber
l_5&=(1,3,2,4)\\ \nonumber
l_6&=(4,2,3,1).
\end{align}
Equation (\ref{eq:3DCalculus}) then tells us that (noting that $n_i=4$ for each $i$)
\begin{align}
A=&\frac{1}{18}\sum_{i=1}^6\sum_{k=1}^{4}F_{ij_{ik}j_{i(k+1)}}\\
=&\frac{1}{6}(F_{146}+F_{163}+F_{135}+F_{154} \label{eq:CubeCalculus}\\ \nonumber
&\ \ +F_{253}+F_{236}+F_{264}+F_{245})\\
=&[1235]-[1236]-[1245]+[1246] \label{eq:CubeTriangulation} \\ 
=&V[12][34][56].
\end{align}
The triangulation (\ref{eq:CubeTriangulation}) came about via one particular choice of simplifying (\ref{eq:CubeCalculus}) using (\ref{eq:main3DIdentity}).  The expression (\ref{eq:CubeCalculus}) encodes all possible triangulations by making different choices of simplification using (\ref{eq:main3DIdentity}).  Moreover, we see that the $V[\cdot\cdot][\cdot\cdot][\cdot\cdot]$ expressions are interpreted as volumes of ``3-D quadrilaterals'' just as the two-dimensional $V[\cdot\cdot][\cdot\cdot]$ objects are interpreted as areas of quadrilaterals.

\section{Vertex $d$-Polytopes}\label{app:DPolytopes}
In this appendix we will briefly describe how vertex $d$-polytopes are defined.  Let $d$ be the dimension of the polytope that we want to define and let $S=\{1, ..., N\}$ be a set of $N\geq d+1$ distinct boundaries, i.e., a set $\{Z_1^{\alpha}, ..., Z_{N}^{\alpha}\}\subset \mathbb{CP}^d$. A vertex in $d$ dimensions is specified uniquely by the intersection of $d$ distinct boundaries and is denoted by $\{i_1, ..., i_d\}$ with $\{i_k\in S\}$ pairwise distinct. A line is specified uniquely by the intersection of $(d-1)$ boundaries, and so an oriented edge is denoted by $[j(i_1, ..., i_{(d-1)})k]$ with $i_1, ..., i_{(d-1)}, j,k\in S$ and with $\{i_1, ..., i_{(d-1)}\}$ pairwise distinct, and denotes the instruction $\{j,i_1, ..., i_{(d-1)}\}\rightarrow \{k, i_1, ..., i_{(d-1)}\}$ along the line $i_1\cap ...\cap i_{(d-1)}$. We place the usual additive structure on formal sums of oriented edges.\\
\indent  Our $d$-polytopes will then be specified by $N\choose d-2$ cyclic lists \begin{equation}
\{l_{i_1, ..., i_{(d-2)}}=(j_{i_1, ..., i_{(d-2)},1}, ...,j_{i_1, ..., i_{(d-2)},n_{i_1, ..., i_{(d-2)}}})\}
\end{equation}
with $i_1, ..., i_{(d-2)}\in S$ and each $j_{i_1, ..., i_{(d-2)},l}\in S-\{i_1, ..., i_{(d-2)}\}$.  Here, $n_{i_1, ..., i_{(d-2)}}$ is just the length of the list $l_{i_1, ..., i_{(d-2)}}$.  The edge set $E_{i_1, ..., i_{(d-2)}}$ derived from the cyclic list $l_{i_1, ..., i_{(d-2)}}$ is defined to be
\begin{equation}
E_{i_1, ..., i_{(d-2)}}=\sum_{l=1}^{n_{i_1, ..., i_{(d-2)}}}[j_{i_1, ..., i_{(d-2)},(l-1)}(j_{i_1, ..., i_{(d-2)},l},i_1, ..., i_{(d-2)})j_{i_1, ..., i_{(d-2)},(l+1)}],
\end{equation}
where the sum is cyclic in the usual sense, and the edge set $E_{i_1, ..., i_{(d-2)};s}$ derived from the cyclic list $l_{i_1, ..., i_{(d-2)}}$ with respect to $s\in S$ is defined as
\begin{align}
E_{i_1, ..., i_{(d-2)};s}&=\sum_{l|j_{i_1, ..., i_{(d-2)},l}=s}^{n_{i_1, ..., i_{(d-2)}}}[j_{i_1, ..., i_{(d-2)},(l-1)}(i_1, ..., i_{(d-2)},j_{i_1, ..., i_{(d-2)},l})j_{i_1, ..., i_{(d-2)},(l+1)}]\\ \nonumber
&=\sum_{l|j_{i_1, ..., i_{(d-2)},l}=s}^{n_{i_1, ..., i_{(d-2)}}}[j_{i_1, ..., i_{(d-2)},(l-1)}(i_1, ..., i_{(d-2)},s)j_{i_1, ..., i_{(d-2)},(l+1)}].
\end{align}
We can now easily generalize our definitions of vertex 2-, 3-, and 4-polytopes to a $d$-polytope for any $d\geq 2$.
\begin{defn}
A \textbf{vertex $d$-polytope} with $d\geq 2$ is equivalent to the following data:\\
\indent i) A set $S=\{1, ..., N\}$ of $N\geq d+1$ distinct boundaries,\\
\indent ii) A collection of $N\choose d-2$ cyclic lists 
\[
\{l_{i_1, ..., i_{(d-2)}}=(j_{i_1, ..., i_{(d-2)},1}, ...,j_{i_1, ..., i_{(d-2)},n_{i_1, ..., i_{(d-2)}}})\}
\]
as defined above, such that for any $i_1, ..., i_{d-1}\in S$, 
\[
E_{i_1, ..., i_{(d-2)};i_{(d-1)}}=(-1)^{|\sigma|}E_{\sigma(i_1, ..., i_{(d-2)};i_{(d-1)})}
\]
where $\sigma\in S_{(d-1)}$ is any permutation of $(d-1)$ objects.
\end{defn}

We note that this definition makes it clear that the restriction of a $d$-polytope to any number of boundaries (say, $p$ distinct boundaries) gives a $(d-p)$-polytope.  Namely, if we are given a $d$-polytope, then the $(d-1)$-polytope $P_I$ obtained by restricting to the $I^{th}$ face is indeed a polytope, since we then have
\begin{equation}
E_{I, ..., i_{(d-3)};i_{(d-2)}}=(-1)^{|\sigma|}E_{I\sigma(i_1, ..., i_{(d-3)};i_{(d-2)})}.
\end{equation}
The analogous statement can be said after restricting to the boundaries $I_1, ..., I_p$ with $p\leq d-2$.\\


\end{document}